\documentclass{mn2e}

\usepackage{amsfonts}
\usepackage{amsmath}
\usepackage{graphicx}

\title[A SEARCH FOR THE MOST MASSIVE GALAXIES. II.]
      {A Search for the Most Massive Galaxies. II. \\
       Structure, Environment and Formation}

\author[Bernardi et al.]
{M. Bernardi$^1$, J. B. Hyde$^1$, A. Fritz$^1$, R. K. Sheth$^1$,
 K. Gebhardt$^2$ \& R. C. Nichol$^3$\\
 $1$ Department of Physics and Astronomy,
     University of Pennsylvania, 209 South 33rd Str., Philadelphia, 
     PA 19104, USA\\
 $2$ Astronomy Department, University of Texas, 1 University Station, 
     C1400, Austin, TX 78712, USA  \\
 $3$ Institute of Cosmology and Gravitation,
     Mercantile House, Hampshire Terrace,
     Univ. of Portsmouth, Portsmouth, PO1 2EG, UK 
}

\begin{document}
\pagerange{\pageref{firstpage}--\pageref{lastpage}}

\maketitle

\label{firstpage}

\begin{abstract}
We study a sample of 43 early-type galaxies, selected from the 
Sloan Digital Sky Survey (SDSS) because they appeared to have 
velocity dispersions of $\sigma\ge$350~kms$^{-1}$.  
High-resolution photometry in the SDSS $i$ passband using the 
High-Resolution Channel of the Advanced Camera for Surveys 
on board the Hubble Space Telescope shows that just less than 
half of the sample is made up of superpositions of two or three 
galaxies, so the reported velocity dispersion is incorrect.  
The other half of the sample is made up of single objects with 
genuinely large velocity dispersions.  None of these objects has 
$\sigma$ larger than $426\pm 30$~km~s$^{-1}$.  
These objects define rather different size-, mass- and 
density-luminosity relations than the bulk of the early-type galaxy 
population:  for their luminosities, they are the smallest, most 
massive and densest galaxies in the Universe.  
Although the slopes of the scaling relations they define are 
rather different from those of the bulk of the population, they 
lie approximately parallel to those of the bulk {\em at fixed $\sigma$}.  
This suggests that these objects are simply the large-$\sigma$ 
extremes of the early-type population -- they are not otherwise 
unusual.  
These objects appear to be of two distinct types:
the less luminous ($M_r>-23$) objects are rather flattened,  
and their properties suggest some amount of rotational support.  
While this may complicate interpretation of the SDSS velocity 
dispersion estimate, and hence estimates of their dynamical mass 
and density, we argue that these objects are extremely dense for 
their luminosities, suggesting merger histories with abnormally 
large amounts of gaseous dissipation.
The more luminous objects ($M_r<-23$) tend to be round and to lie 
in or at the centers of clusters.  Their circular isophotes, large 
velocity dispersions, and environments are consistent with the 
hypothesis that they are BCGs.  Models in which BCGs form from 
predominantly radial mergers having little angular momentum predict 
that they should be prolate.  If viewed along the major axis, such 
objects would appear to have abnormally large velocity dispersions 
for their sizes, and to be abnormally round for their luminosities.  
This is true of the objects in our sample once we account for the fact 
that the most luminous galaxies ($M_r<-23.5$), and BCGs, become slightly 
less round with increasing luminosity.  Thus, the shapes of the most 
luminous galaxies suggest that they formed from radial mergers, and 
the shapes of the most luminous objects in our big-$\sigma$ sample 
suggest that they are the densest of these objects, viewed along the 
major axis.  
\end{abstract}

\begin{keywords}
galaxies: elliptical and lenticular, cD
--- galaxies: evolution
--- galaxies: kinematics and dynamics
--- galaxies: structure
--- galaxies: stellar content  
\end{keywords}

\section{Introduction}
The most massive galaxies may place interesting constraints on models 
of galaxy formation (e.g. De Lucia et al. 2006; Almeida et al. 2007).  
But which observable one should use as a proxy for mass is debatable.  
If luminosity is a good proxy, then the Brightest Cluster Galaxies 
should be the most massive galaxies; this has led to considerable 
interest in their properties (Scott 1957; Sandage 1976; 
Thuan \& Romanishin 1981; 
Malumuth \& Kirschner 1981; Hoessel et al. 1987; Schombert 1987, 1988;
Oegerle \& Hoessel 1991; Lauer \& Postman 1994; Postman \& Lauer 1995; 
Crawford et al. 1999; Laine et al. 2003; Lauer et al. 2007; 
Bernardi et al. 2007a).  

\begin{figure*}
 \centering
 \includegraphics[scale=1.33]{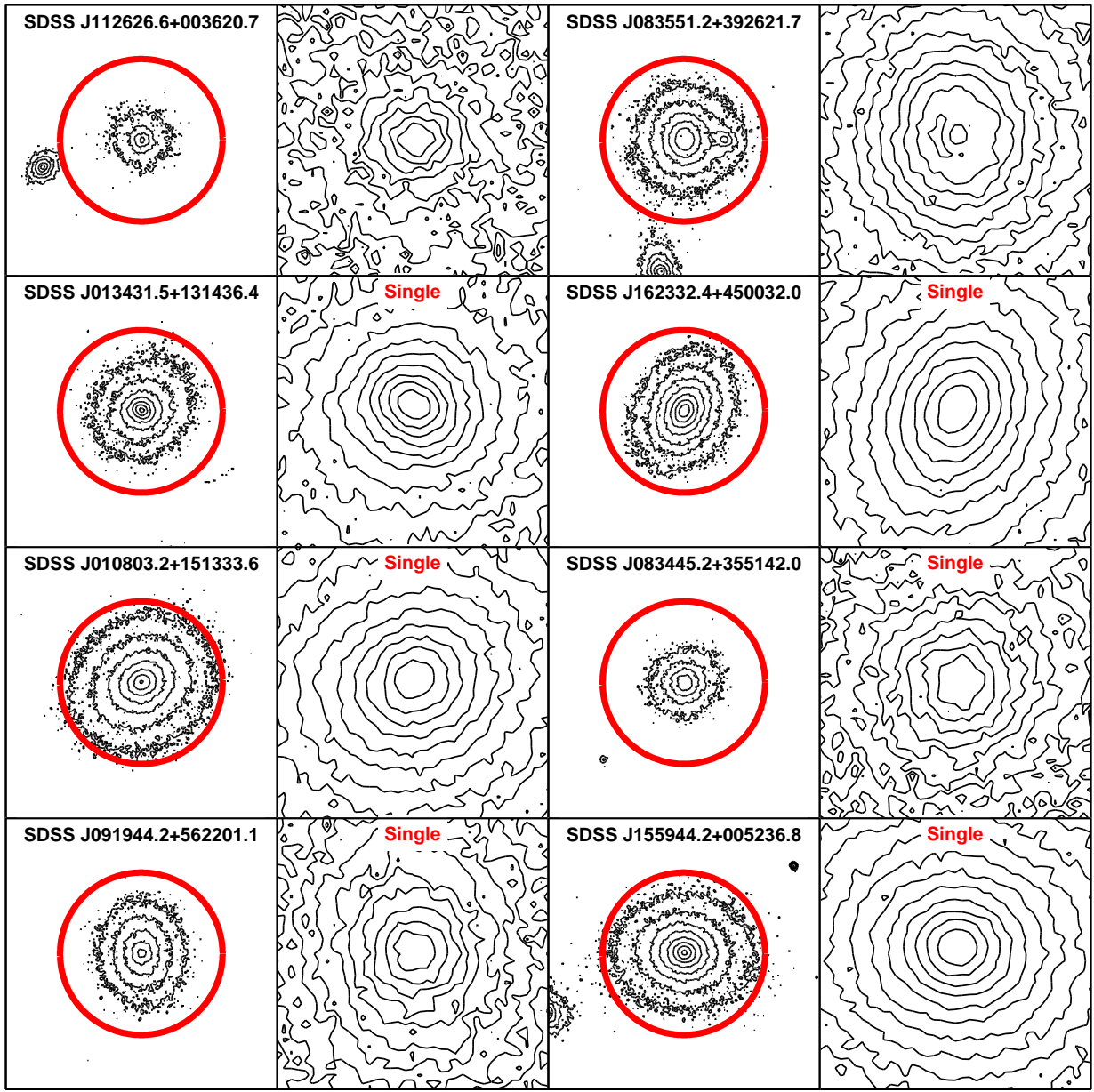}
 \caption{Surface brightness isophotes of the objects in our sample, 
          which provide the basis for determining which objects are 
          singles.  The two panels for each object show a larger 
          $5\times 5$~arcsec region with logarithmically spaced 
          isophotes, and a $1\times 1$~arcsec region with linearly 
          spaced isophotes.  Thick solid circle shows the size of 
          the SDSS fiber; structure within this circle is likely to 
          have contributed to the estimated velocity dispersion, 
          whereas structures outside it have not.}
 \label{A1}
\end{figure*}


\begin{figure*}
 \centering
 \includegraphics[scale=1.33]{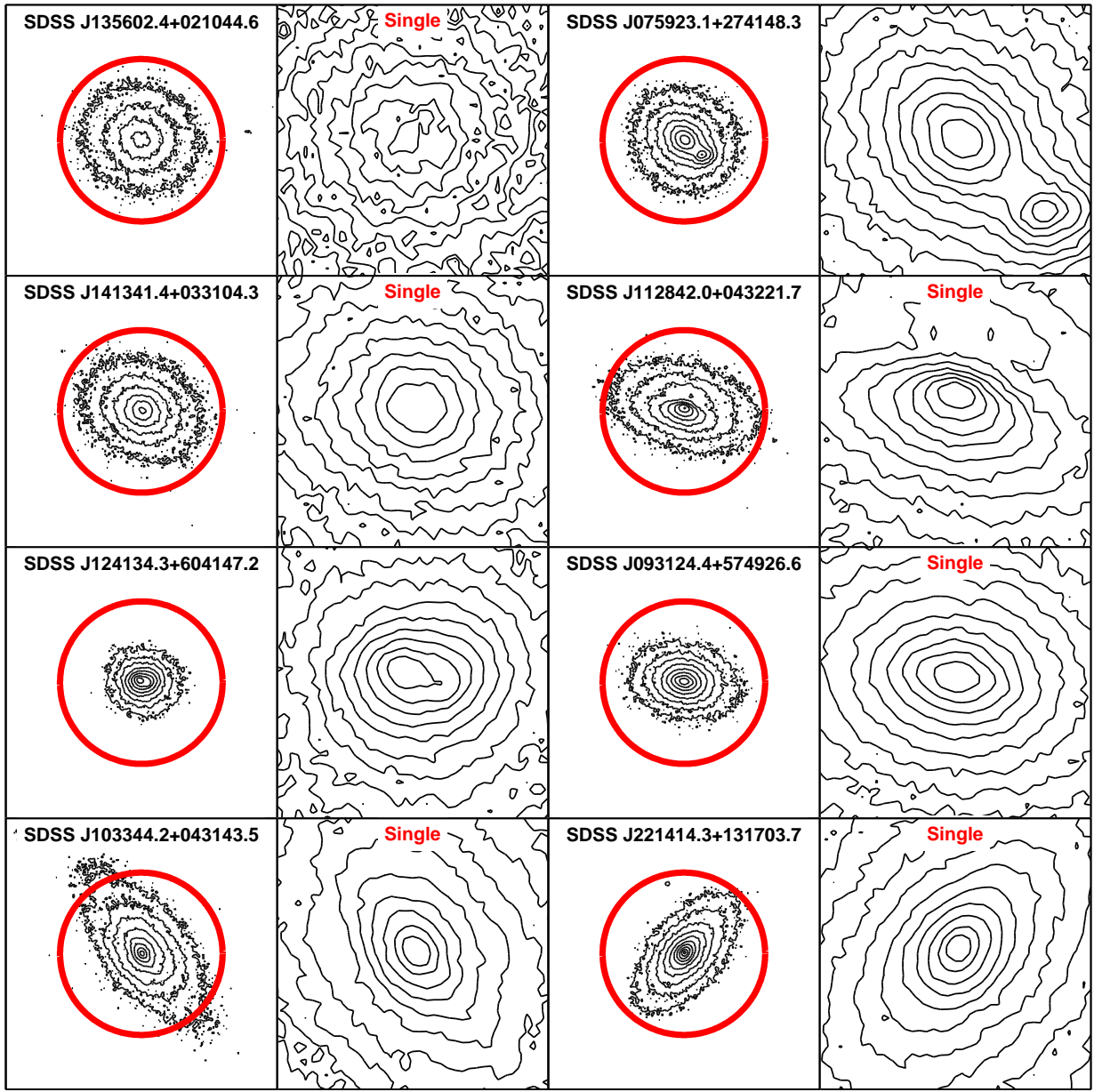}
 \caption{Continued from previous figure.  The assymetry in in the image
          which is second from top on the right is due to dust.}
 \label{A2}
\end{figure*}


\begin{figure*}
 \centering
 \includegraphics[scale=1.33]{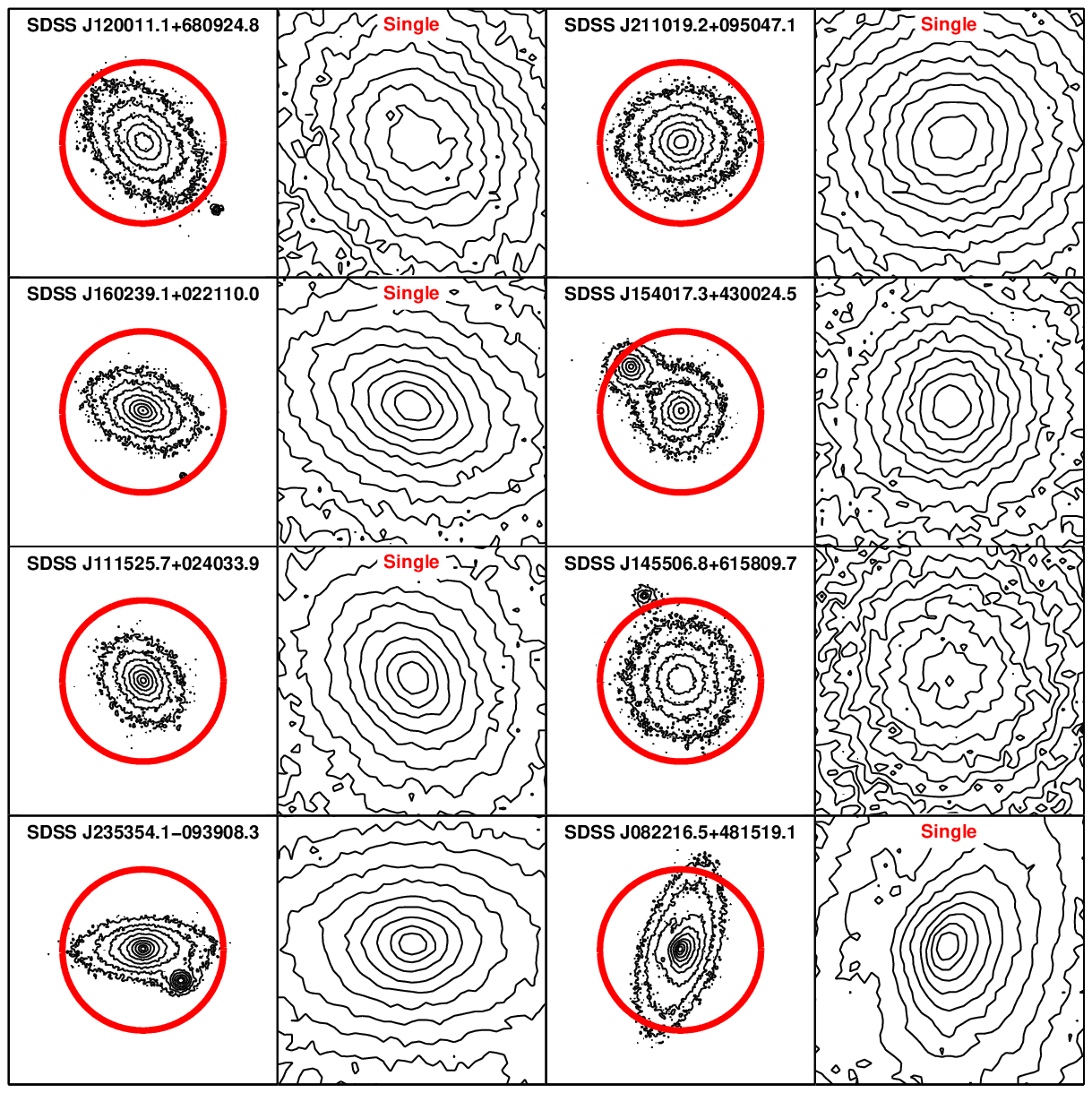}
 \caption{Continued from previous figure.}
 \label{A3}
\end{figure*}


\begin{figure*}
 \centering
 \includegraphics[scale=1.33]{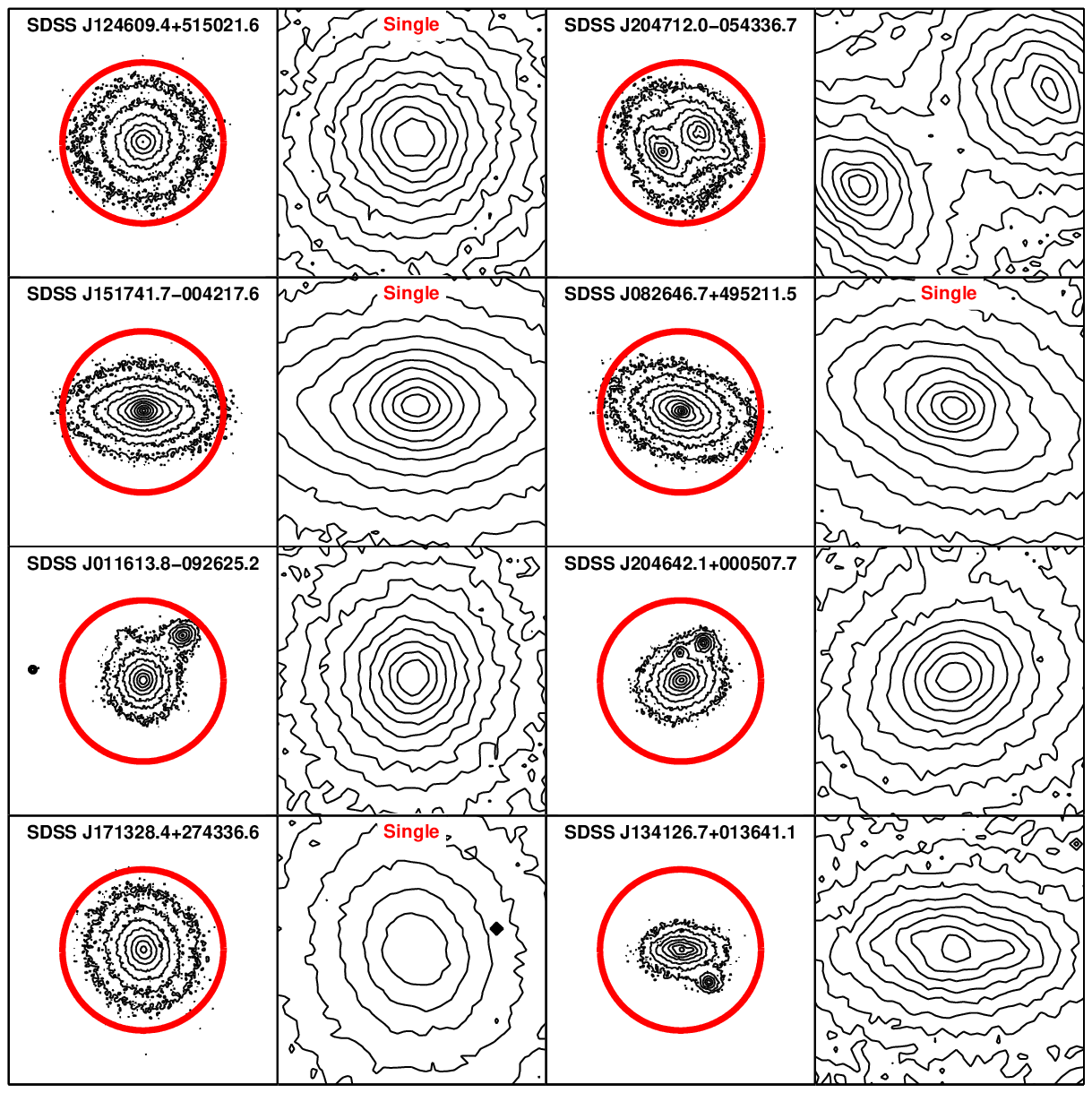}
 \caption{Continued from previous figure.}
 \label{A4}
\end{figure*}


\begin{figure*}
 \centering
 \includegraphics[scale=1.33]{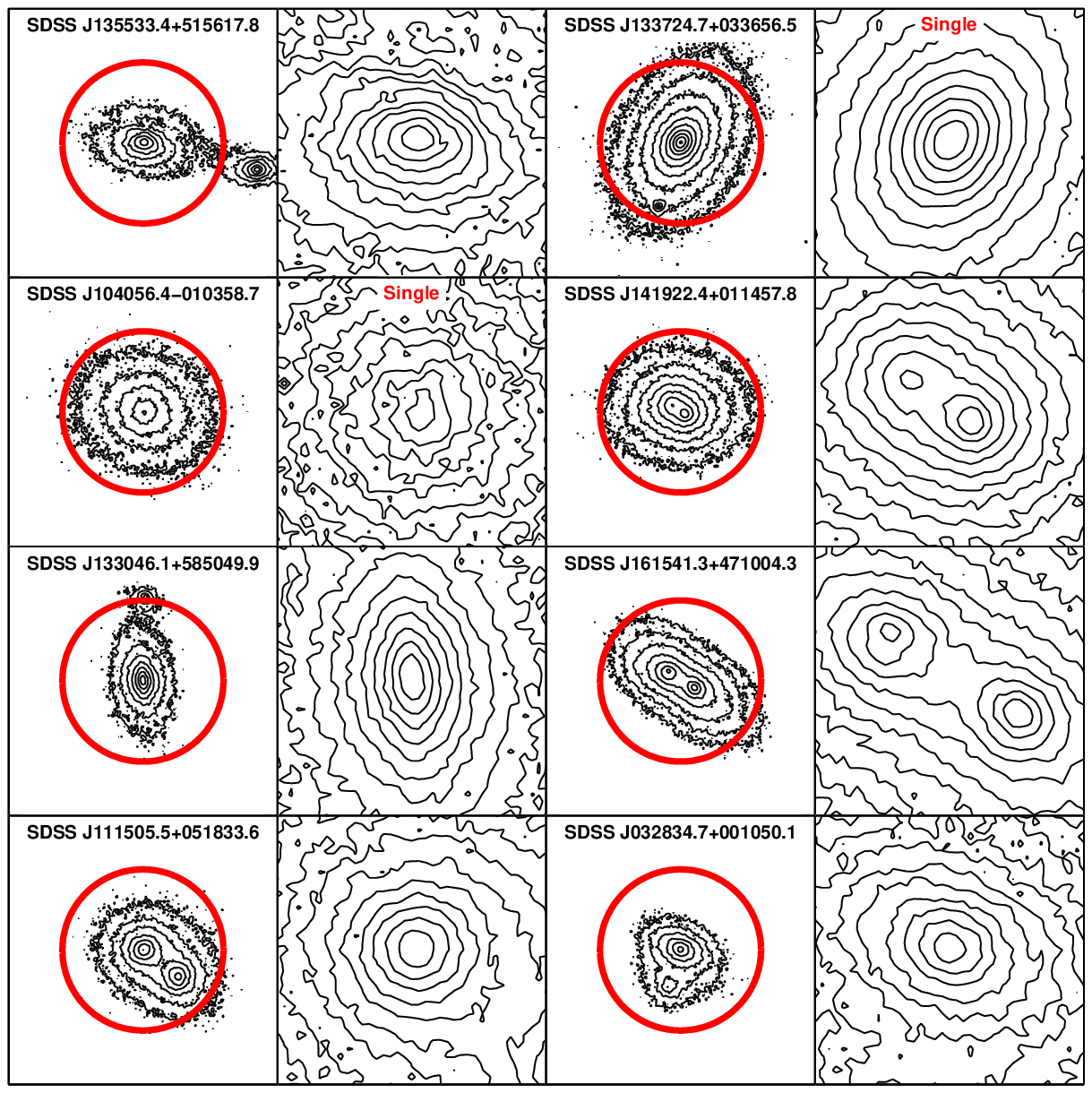}
 \caption{Continued from previous figure.}
 \label{A5}
\end{figure*}


\begin{figure*}
 \centering
 \includegraphics[scale=1.33]{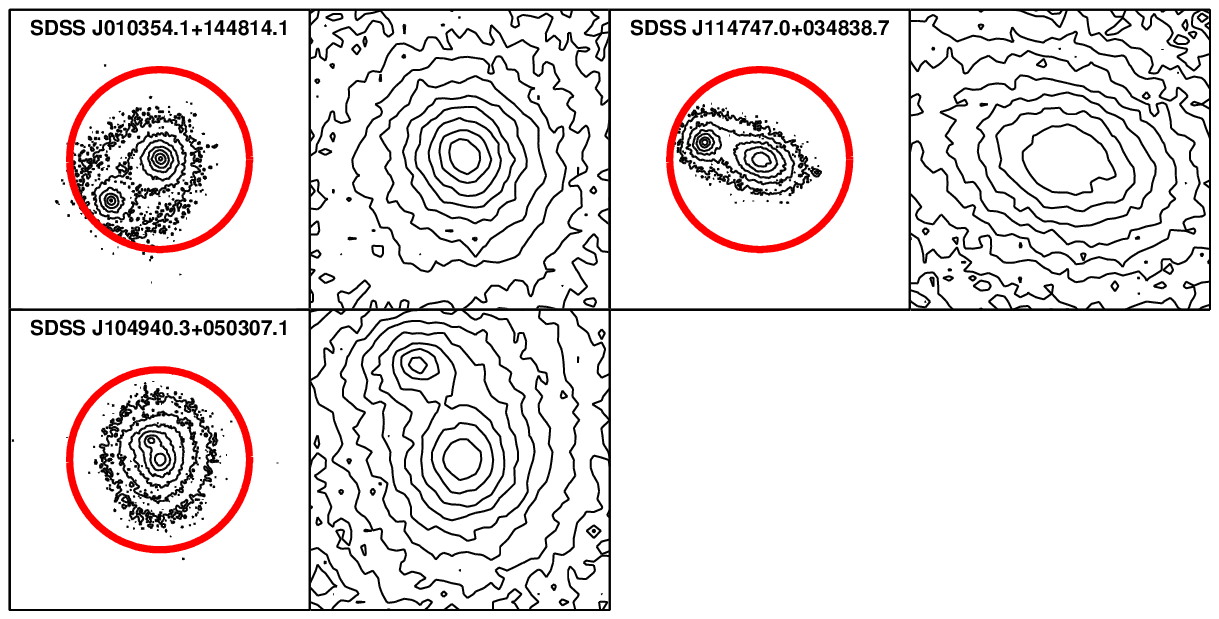}
 \vspace{-8cm}
 \caption{Continued from previous figure.}
 \label{A6}
\end{figure*}

However, velocity dispersion is sometimes used as a surrogate for 
mass (the virial theorem has mass $\propto R\sigma^2$), so it is 
interesting to ask if a sample selected on the basis of velocity 
dispersion also contains the most massive galaxies.  
Such a sample is also interesting in view of the fact that black 
hole mass correlates strongly and tightly with velocity dispersion 
(Ferrarese \& Merritt 2000; Gebhardt et al. 2000), so galaxies 
with large $\sigma$ are expected to host the most massive black 
holes.  

With this in mind, Bernardi et al. (2006) 
culled a sample of $\sim 100$ objects with $\sigma > 350$~km~s$^{-1}$ 
from the First Data Release of the Sload Digital Sky Survey 
(SDSS DR1; Abazajian et al. 2003).  The total area from which these 
objects were selected is about 2000~deg$^2$; for a spatially flat 
Universe with $\Omega_m=0.3$ and Hubble constant $H_0 = 70$~km~s$^{-1}$, 
which we assume in what follows, this corresponds to a comoving 
volume of about $3.34\times 10^8$~Mpc$^3$ out to $z=0.3$.

Some of these objects turned out to be objects in superposition, 
evidence for which came primarily from the spectra (see Bernardi et al. 2006
for details).  
A random subset of the others, 43 objects in all, was 
observed with the High Resolution Camera of the Advanced 
Camera System on board the Hubble Space Telescope (HST-ACS HRC).  
On the basis of this high-resolution imaging, we have been able 
to separate out the true singles from those which are objects 
in superposition.  As a result, we are now able to study the 
properties of objects with large $\sigma$.  

We describe the HST observations, our identification of the truly 
single objects, our classification of their isophotal shapes, and 
how we combine the HST imaging with SDSS data to determine the 
environments of these objects in Section~\ref{hst}.  
The isophotal shapes of these objects are compared with those 
of BCGs in  Section~\ref{shapes}.  
Various scaling relations, size-luminosity, the fundamental plane, 
etc. are presented in Section~\ref{scaling}.  While these 
relations are different from those defined by the bulk of the 
early-type galaxy population, we show that they can actually be 
derived from the early-type scaling relations simply by studying 
what these relations look like at fixed velocity dispersion.  
These analyses show that our sample appears to contain two distinct 
types of objects:  the more luminous objects tend to be round, have 
large sizes as well as velocity dispersions, and tend to be in crowded 
fields.  Indeed, they are rounder than other objects of similar 
luminosities, so it is possible that they are prolate, and viewed 
along the long axis.  
The less luminous objects are not round, have abnormally small 
sizes and are not necessarily in crowded fields; even if rotational 
motions have compromised the SDSS velocity dispersion estimates, 
these objects may be amongst the densest galaxies for their 
luminosities.  
A final section summarizes our findings and discusses some 
implications of the bimodal distribution we have found, as 
well as of the fact that no galaxy appears to have a velocity 
dispersion larger than $426\pm30$~km~s$^{-1}$.  

In a companion paper (Hyde et al. 2008), we study the surface 
brightness profiles of the objects classified as singles in much 
more detail, placing them in the context of other HST-based studies 
of early-type galaxies (e.g. Laine et al. 2003; Ferrarese et al. 2006; 
Lauer et al. 2007).

\section{The data}\label{hst}

\subsection{HST Observations}
The analysis below is based on observations taken between August 2004 
and June 2005 as part of a Cycle 13 Snapshot Survey.
The target list for the Snapshot Survey contained 70 objects, all of 
which were selected from the SDSS as described in Paper I; 
i.e., they all had reported velocity dispersions larger than 
350~km~s$^{-1}$, and none were identified from the imaging, the line 
profiles or the cross-correlation analyses described in Paper I as 
being multiple.  Of these 70 objects, 43 were observed.

The following observing sequence was adopted for each target galaxy: \\
-The center of each galaxy was positioned close to the HRC aperture.\\
-A total integration time of 1200~s in the F775W (which resembles 
SDSS $i$) filter, split into four exposures of 300~s, was used.\\
-Each visit used the line-dither pattern to allow removal of most 
cosmic ray events.\\
Data reduction details and analysis of the photometric profiles are 
presented in Hyde et al. (2008).

\subsection{Identifying the singles}
In the present context, the great virtue of HST is its angular 
resolution ($\sim 0.025$~arcsec/pixel): 
at the median redshift of our sample, $z=0.24$, 
scales down to about 95~pc are resolved.  (This scale, for SDSS 
imaging, is about sixty times larger.)  
Therefore, a simple visual inspection of the F775W images allowed 
a classification into three groups:\\
  (i) 23 targets showed no evidence for superposition - we 
      call these single galaxies; \\
 (ii) 15 targets appear to be superpositions of two galaxies - we 
      call these doubles;\\
(iii) 5 targets show evidence for more than two components - we 
      call these multiples.

Figures~\ref{A1}--\ref{A6} show density contours (isophotes) 
obtained from the ACS F775W images of these objects.  Two panels 
are shown for each object.  The panels on the left show 
fields-of-view that are approximately $5 \times 5$~arcsec$^2$ 
(North is up, East is to the left), and the contours are linearly 
spaced in magnitude (logarithmically spaced in flux).  
The solid circle shows the angular size of the SDSS fiber (3 arcsec 
in diameter).
This is the scale within which the velocity dispersion was measured 
(this measured value is aperture corrected to $R_e/8$ - see 
Bernardi et al. 2006 for details).  
Light from objects outside the circle is very unlikely to affect 
the velocity dispersion and absorption line-index measurements.
The panels on the right show the central arcsecond, and the contours 
are linearly spaced in flux.  In these panels, we specify if the 
object was classified as a single.  
These figures illustrate that the classification into singles, 
doubles and multiples is relatively straightforward.
Table~\ref{tab:classify} provides details of these classifications.

\begin{figure}
 \begin{center}
 \includegraphics[scale=0.45]{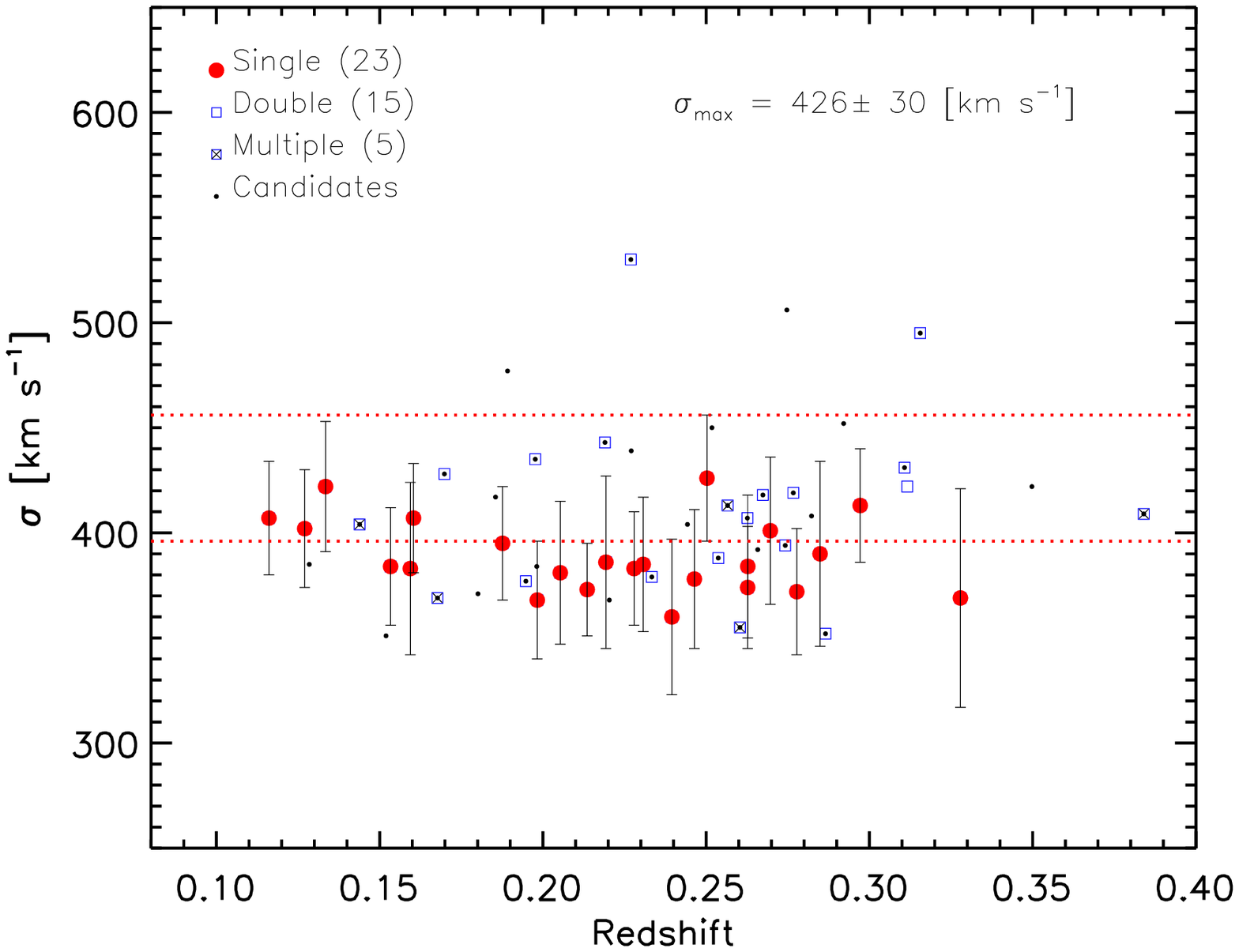}
 \caption{Velocity dispersions of early-type galaxies in our sample
 sample as a function of redshift. Circles represent the singles,
 squares show the doubles, and squares with an additional cross 
 show objects which are superpositions of more than two components.
 Small filled squares show objects which Bernardi et al. (2006) 
 thought were likely to be massive single galaxies.  We only show 
 measurement uncertainties on the singles; uncertainties on the 
 other objects are similar, although the systematic error is, of 
 course, substantially larger.}
 \label{sigz}
 \end{center}
\end{figure}

\begin{table}
 \caption[]{The 43 objects with velocity dispersions larger than 
            350~km~s$^{-1}$ from Bernardi et al. (2006) which 
            have HST images.  ID$_{\rm S}$ identifies the single 
            objects whose properties are given in Table~2.   
            N$_{\rm objs}$ gives the number of objects inside a 
            circle of 3~arcsec\ in diameter.
            Env = 0,1 means low-, high-density (see text for details).
            Prof = c,p,d means core, power-law, obvious presence of dust 
            (from Hyde et al. 2008).\\}
\centering
\begin{tabular}{ccccc}
 \hline &&\\
  Name & ID$_{\rm S}$ & N$_{\rm objs}$ & Env & Prof\\
\hline &&\\
SDSS J112626.6$+$003620.7 &-- & 2  &  0  & -- \\
SDSS J083551.2$+$392621.7 &-- & 3  &  1  & -- \\
SDSS J013431.5$+$131436.4 & 1 & 1  &  0  & p  \\
SDSS J162332.4$+$450032.0 & 2 & 1  &  1  & c  \\
SDSS J010803.2$+$151333.6 & 3 & 1  &  0  & c  \\
SDSS J083445.2$+$355142.0 & 4 & 1  &  1  & c  \\
SDSS J091944.2$+$562201.1 & 5 & 1  &  1  & c  \\
SDSS J155944.2$+$005236.8 & 6 & 1  &  0  & c  \\
SDSS J135602.4$+$021044.6 & 7 & 1  &  1  & c  \\
SDSS J075923.1$+$274148.3 &-- & 2  &  1  & -- \\
SDSS J141341.4$+$033104.3 & 8 & 1  &  0  & c  \\
SDSS J112842.0$+$043221.7 & 9 & 1  &  0  & p,d  \\
SDSS J124134.3$+$604147.2 &-- & 2  &  0  & -- \\
SDSS J093124.4$+$574926.6 &10 & 1  &  0  & p  \\
SDSS J103344.2$+$043143.5 &11 & 1  &  0  & p,d  \\
SDSS J221414.3$+$131703.7 &12 & 1  &  0  & p  \\
SDSS J120011.1$+$680924.8 &13 & 1  &  1  & c  \\
SDSS J211019.2$+$095047.1 &14 & 1  &  1  & c  \\
SDSS J160239.1$+$022110.0 &15 & 1  &  1  & p  \\
SDSS J154017.3$+$430024.5 &-- & 2  &  1  & -- \\
SDSS J111525.7$+$024033.9 &16 & 1  &  0  & p  \\
SDSS J145506.8$+$615809.7 &-- & 2  &  1  & -- \\
SDSS J235354.1$-$093908.3 &-- & 2  &  0  & -- \\
SDSS J082216.5$+$481519.1 &17 & 1  &  0  & p,d  \\
SDSS J124609.4$+$515021.6 &18 & 1  &  1  & c  \\
SDSS J204712.0$-$054336.7 &-- & 3  &  1  & -- \\
SDSS J151741.7$-$004217.6 &19 & 1  &  1  & p  \\
SDSS J082646.7$+$495211.5 &20 & 1  &  0  & p,d  \\
SDSS J011613.8$-$092625.2 &-- & 2  &  0  & -- \\
SDSS J204642.1$+$000507.7 &-- & 3  &  1  & -- \\
SDSS J171328.4$+$274336.6 &21 & 1  &  1  & c  \\
SDSS J134126.7$+$013641.1 &-- & 3  &  0  & -- \\
SDSS J135533.4$+$515617.8 &-- & 2  &  0  & -- \\
SDSS J133724.7$+$033656.5 &22 & 1  &  1  & c  \\
SDSS J104056.4$-$010358.7 &23 & 1  &  1  & c  \\
SDSS J141922.4$+$011457.8 &-- & 2  &  1  & -- \\
SDSS J133046.1$+$585049.9 &-- & 2  &  0  & -- \\
SDSS J161541.3$+$471004.3 &-- & 2  &  1  & -- \\
SDSS J111505.5$+$051833.6 &-- & 2  &  0  & -- \\
SDSS J032834.7$+$001050.1 &-- & 2  &  0  & -- \\
SDSS J010354.1$+$144814.1 &-- & 2  &  0  & -- \\
SDSS J114747.0$+$034838.7 &-- & 2  &  0  & -- \\
SDSS J104940.3$+$050307.1 &-- & 2  &  1  & -- \\
\hline &&\\
\end{tabular}
\label{tab:classify} 
\end{table}

Finally, note that a few of the objects in this sample 
(SDSS J103344.2$+$043143.5, SDSS J112842.0+043221.7,  
SDSS J082646.7+495211.5 and SDSS J082216.5+481519.1)
show clear evidence for dust; this is studied in more detail 
in Hyde et al. (2008).

\begin{figure}
 \begin{center}
 \includegraphics[scale=0.4]{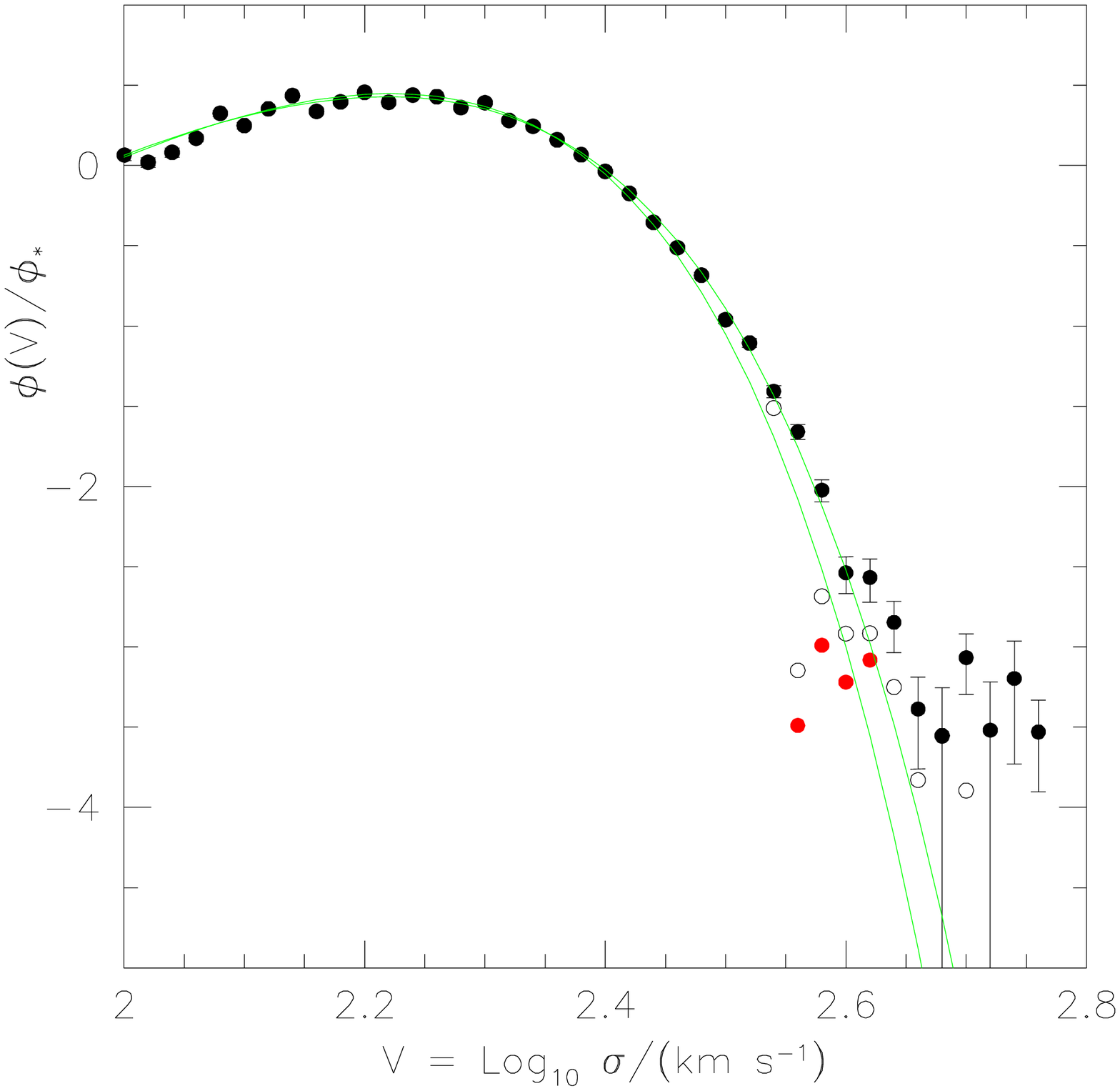}
 \caption{Normalized distribution of SDSS early-type velocity dispersions before 
          and after removing the contribution from objects which 
          are superpositions.  Filled circles with error bars show 
          the measured distribution with no correction (other than 
          to account for the apparent magnitude limit by weighting 
          each object by $1/V_{\rm max}$); 
          open symbols show the result of removing superpositions 
          on the basis of SDSS imaging or spectra (e.g. following 
          Bernardi et al. 2006) - since this does not find all the 
          superpositions, this will overestimate the true abundance; 
          and filled circles without error bars show the result of 
          assuming that the only singles are those identified by 
          HST - since this incorrectly assumes that all the others 
          are superpositions, it underestimates the true abundance.  
          Smooth curves show fits to the functional form given by 
          Sheth et al. (2003); the curve which predicts lower 
          abundances at large $\sigma$ is an estimate of the 
          intrinsic distribution.  Convolving it with measurement 
          errors yields the other curve.}
 \label{phiv}
 \end{center}
\end{figure}

\subsection{Parameters from SDSS images and spectra}\label{photocorr}
The SDSS imaging reductions are known to suffer from sky subtraction 
problems, particularly for large objects. To address this problem,
for the sample of large velocity dispersion galaxies, 
we have recomputed the photometric parameters 
(i.e. deVaucouleur magnitudes, sizes, $b/a$ and model color) from the 
SDSS r-band images using {\it GalMorph} (Hyde et al. 2008).
The photometric parameters of the entire SDSS early-type sample were 
corrected following equations~(1-4) of Hyde \& Bernardi (2008).   
The spectroscopic parameters, velocity dispersions and Mg$_2$ 
index-strengths for the objects with $\sigma>350$~km~s$^{-1}$ are 
from Bernardi et al. (2006); those for the rest of the early-type 
sample are from DR6 (since these do not suffer from the bias at 
small $\sigma$ that is present in previous SDSS data releases -- see 
DR6 documentation and Bernardi 2007). Table~2 lists the parameters used 
in this work.  A more extensive analysis of the HST (rather than SDSS) 
surface brightness profiles of these objects is provided by 
Hyde et al. (2008).

\subsection{The environments of these objects}
In most of the analysis which follows, the environment of these 
objects is irrelevant.  
However, whether or not the objects with the largest velocity 
dispersions are preferentially found in dense environments is an 
interesting question.  This is because, in hierarchical models, 
the most massive halos are predicted to populate the densest regions 
(Mo \& White 1996; Sheth \& Tormen 2002), a trend for which there is 
good observational evidence (e.g. Abbas \& Sheth 2007).  
Massive halos are expected to host the most massive galaxies 
(e.g. De Lucia et al. 2006; Almeida et al. 2007), so if the largest 
velocity dispersions reflect large masses, then one expects a 
correlation between $\sigma$ and environment.  
One might also wish to check if the objects which are superpositions 
are found in particularly crowded fields.  To briefly address these 
questions, we have made a rough estimate of the environment of each 
object as follows.  

From the SDSS DR5, we select fields which are approximately 
1~Mpc $\times$ 1~Mpc centred on each target.  For the most distant 
objects in our sample ($z\sim 0.3$), the fields are about 
$3'\times 3'$, whereas they are about $9'\times 9'$ for the 
closest objects ($z\sim 0.1$).  
The SDSS provides photometric information (colors, photometric 
redshifts) for all the objects in the field which are brighter 
than about $m_r=22.5$.  If spectroscopic information is also 
available (in general, if $m_r$ is brighter than $\sim 17.5$), 
we count the object as a neighbour if it lies within 
3000~km~s$^{-1}$ of the target.  If only photometry is available, 
an object counts as a neighbour if the photometric redshift is 
within 7500~km~s$^{-1}$ of the target.  We then classify the 
targets as being in group or cluster like environments if the 
number of neighbours is greater than 30.  (Changing this 
threshold to 20 makes little difference.)  
We will refer to the objects in these two environments as 
being in {\it high} or {\it low-density}. Table~1 lists our environment
classification for each galaxy and Figures~\ref{ba}-\ref{fig:mg2se}
use this classification to separate galaxies in high and low density
environments.


We have made no effort to account for the SDSS apparent magnitude 
limit, e.g., by only counting neighbours above an absolute, rather 
than apparent, magnitude limit, that is visible across the entire 
survey.  This is in part because the vast majority of our 
high-redshift objects, for which the apparent magnitude limit 
matters most, are classified as being in dense regions anyway.  
In addition, environment does not play a crucial role in 
what follows -- we have included it here to show what is possible.  
For instance, one might prefer to classify environment on the basis 
of distance to the nearest cluster; we leave a more careful analysis 
of the environment for future work.
We note however, that most of the objects with $M_r<-23.5$ are 
likely to be BCGs. 


\subsection{The abundance of galaxies with $\sigma \ge 350$~km~s$^{-1}$}
Figure~\ref{sigz} shows the distribution of velocity dispersions 
as a function of redshift in our sample; different symbol styles 
show the singles, doubles and multiples.  Small filled squares show 
the objects classified in Table~1 of Bernardi et al. (2006) as 
likely to be single galaxies.  
Although the doubles and multiples are not the focus of study in 
this paper, we note that the distribution of doubles is biased 
towards slightly higher redshifts than is the distribution of singles, 
as one might expect.  We cannot make a similar statement about the 
distribution of multiples, because we have so few.   

We turn now to the objects classified as singles.  
All these objects were drawn from the SDSS which is magnitude limited, 
so the mean luminosity at the high redshift end is two magnitudes 
brighter than at the low redshift end (see Figure~\ref{mrzse}).  
Since velocity dispersion and luminosity are correlated 
$\langle\log_{10}\sigma|M_r\rangle \propto 0.1M_r$, the fact that 
Figure~\ref{sigz} shows no trend with redshift suggests that the 
objects at low $z$ have more extreme velocity dispersions for their 
$L$ than the objects at high-$z$.  

In all cases, the velocity dispersions were derived from the SDSS 
spectra (median signal-to-noise S/N=18 per pixel) and aperture 
corrected to $R_e/8$ as described by Bernardi et al. (2006). 
Figure~\ref{phiv} shows the distribution of objects as a function 
of velocity dispersion, in the SDSS, using methods detailed in 
Sheth et al. (2003).  
Symbols with error bars show the measured distribution before 
removing the objects which are superpositions - notice the 
apparent excess at $\sigma \ge 350$~km~s$^{-1}$; open circles 
show the result of removing superpositions identified on the basis 
of SDSS imaging and spectra (following Bernardi et al. 2006), so 
they overestimate the true abundances; 
and filled circles without error bars show the result of 
assuming that the only single objects are those identified by 
our HST analysis - so they underestimate the true abundances.  
This shows that superpositions do affect the large $\sigma$ tail 
significantly, but that, once they have been removed, there is no 
excess of objects at high $\sigma$.  
The single galaxy with the largest aperture corrected velocity 
dispersion has a spectrum with S/N=25, is located in a group
environment at $z=0.25$, and has $\sigma_c=426\pm30$~km~s$^{-1}$.  

It is worth discussing the toe of large 
$\sigma\approx 500$~km~s$^{-1}$ ($\log\sigma\approx 2.7$) 
objects in a little more detail.  
Such a toe is not seen in Figure~1 of Sheth et al. (2003) because 
the SDSS database at the time set a hard upper limit on reported 
dispersions of $\sigma=414$~km~s$^{-1}$.  This is still true:  
objects for which the pipeline would have returned a larger $\sigma$ 
are simply assigned this limiting value, $\sigma = 414$~km~s$^{-1}$, 
in the database.  For this reason, Figure~\ref{phiv} in the current 
draft is based on our own reductions of the objects which SDSS 
reports as having $\sigma > 350$~km~s$^{-1}$ (Bernardi et al. 2006). 
These reductions 
are essentially the same as those of the SDSS, except that we 
use the actual values returned by our analysis pipeline even if 
they exceed $\sigma = 414$~km~s$^{-1}$.  

The Sheth et al. (2003) analysis was based on a small enough 
sample that this limiting value affected only a handful of objects, 
so the toe associated with objects piling up at 414~km~s$^{-1}$ was 
not obvious.  The toe is there in the larger sample, as our 
Figure~\ref{phiv} shows.  Of course, we also argue that this toe is 
almost entirely made up of doubles rather than singles.  
The two smooth curves show an estimate of the intrinsic distribution, 
and the result of convolving it with measurement errors.

\begin{figure}
 \begin{center}
 \includegraphics[scale=0.45]{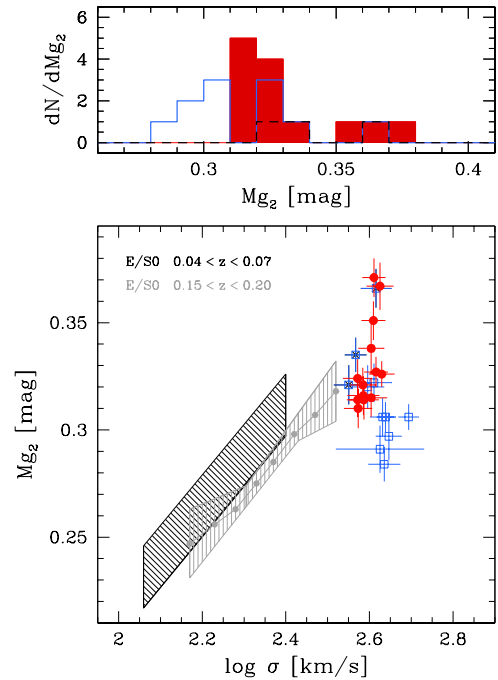}
 \caption{Top:  Distribution of Mg$_2$ index strength in our sample 
  of 43 objects.  Shaded, solid and dashed histograms are for 
  objects classified as singles, doubles and multiples.  
  Bottom: Strength of Mg$_2$ index as function of velocity dispersion. 
  Circles represent the singles, squares show the doubles, 
  and squares with an additional cross show objects which are 
  superpositions of more than two components.
  Note the relatively clear separation between the doubles and the 
  others. The hashed regions show the relation defined by the bulk 
  of the early-type galaxy population at two different redshifts.  
  The big-$\sigma$ singles populate the upper end of the relationship 
  defined by the bulk of the population. }
 \label{fig:mg2sig}
 \end{center}
\end{figure}

\subsection{The Mg$-\sigma$ relation}\label{mgs}
Bernardi et al. (2006) suggested that early-type galaxy scaling 
relations (e.g. $R-L$, $R\sigma^2-L$) could be 
used as a diagnostic for superposition.  The objects in this sample, 
however, all lie sufficiently close to these relations that 
Bernardi et al. were unable to separate the singles from superpostions.  

Bernardi et al. (2006) also argued that correlations between absorption 
line indices and velocity dispersion should provide an efficient way 
of identifying superpositions.  The idea is that superpositions 
which affect the measured $\sigma$ will also affect the measured 
line index strengths.  
For the Mg$_2-\sigma$ relation, for example, objects in close 
superposition are expected to lie below and to the right of the true 
relation.  This was the technique for which the precise place to 
divide between singles and superpositions was least robustly 
determined.  Now that we have the HST imaging to identify the 
superpositions, we can ask how well this technique works.  

Figure~\ref{fig:mg2sig} shows the Mg$_2-\sigma$ relation for our 
sample.  Note the clear separation between singles (filled circles) 
and doubles (open squares):  at a given $\sigma$, doubles do indeed 
have abnormally low Mg$_2$.  For comparison, shaded regions show 
the relations defined by SDSS galaxies in two redshift bins: 
$0.04<z<0.07$ (black) and $0.15<z<0.20$ (grey).  
The singles lie at the large $\sigma$ extremes of these relations, 
a point we will return to later.  
On the other hand, notice that the squares with crosses populate 
the same regions as the filled circles:  evidently, this method 
alone is not effective at isolating objects with more than two 
components.

\section{The shapes of singles: Evidence for two populations?}\label{shapes}
The same galaxy, if viewed along its longest axis, will appear rounder, 
with a larger velocity dispersion and smaller size, than if the line 
of sight is perpendicular to the longest axis.  So it is interesting 
to ask if the large velocity dispersions in our sample are due in part 
to projection effects.  If so, then this may indicate that they formed 
from primarily radial orbits (e.g. Gonz\'alez-Garc\'ia \& van Albada 2005; 
Boylan-Kolchin et al. 2006).

\begin{figure*}
 \begin{center}
 \includegraphics[scale=0.8]{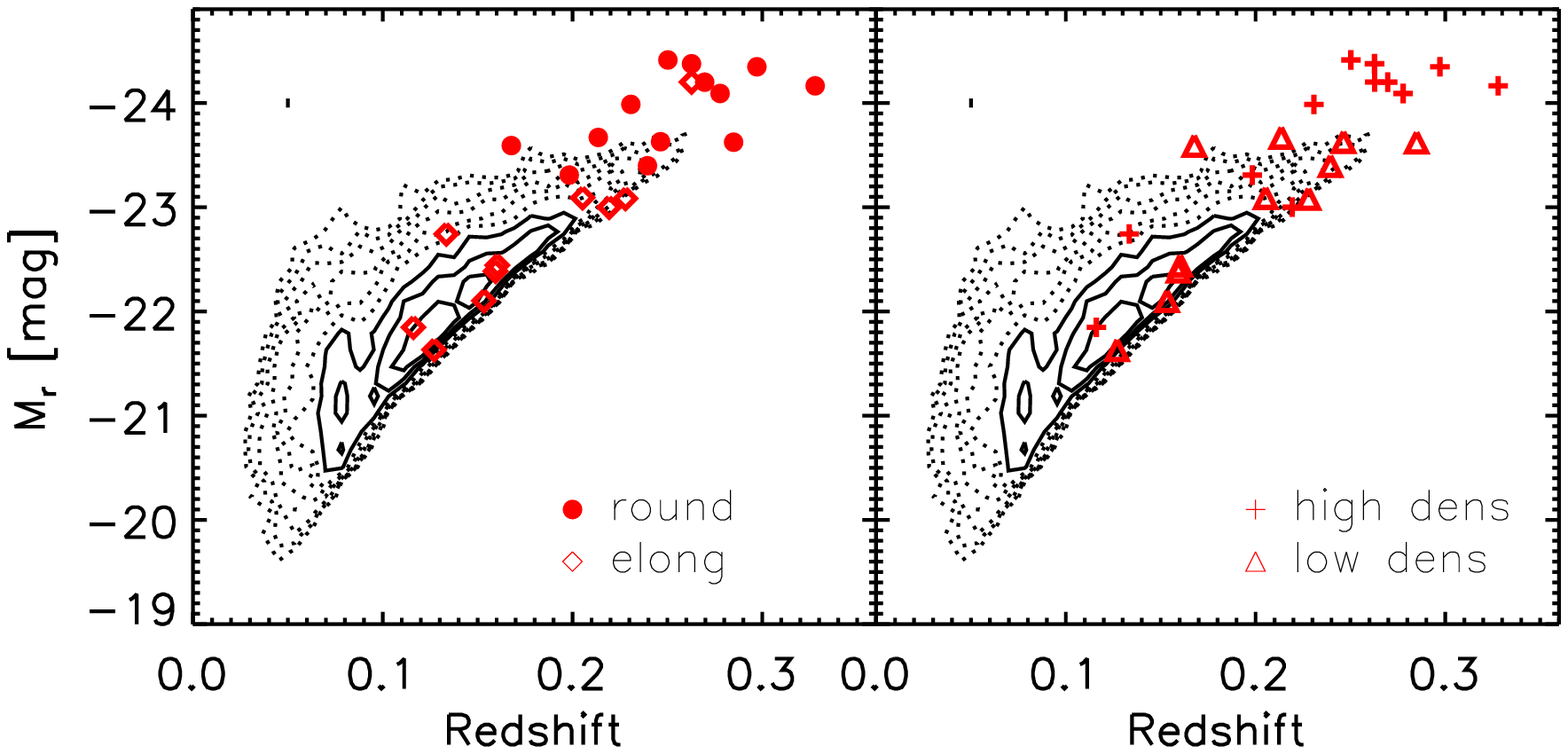}\vspace*{-0.5cm}
 \includegraphics[scale=0.8]{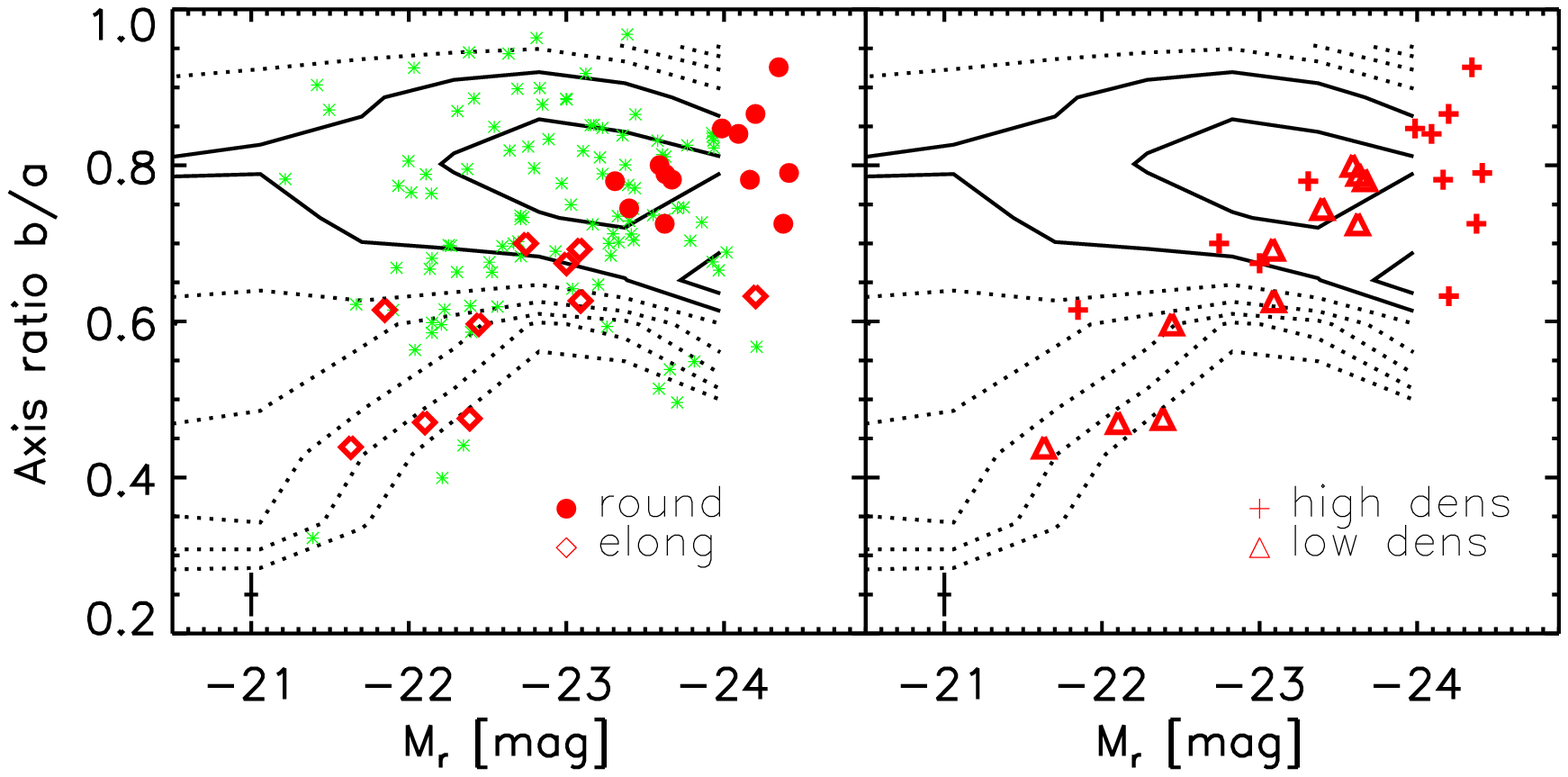}
 \caption{Top:  Distribution of luminosities for galaxies classified as 
      singles.  Countours show the distribution of the early-type galaxy 
      sample of Bernardi et al. (2006). Solid line contours include
      $68\%$ of the objects.
      Filled circles and open diamonds in the panel on the left show 
      objects classified as `round' and `elongated';  
      crosses and triangles in the panel on the right represent galaxies 
      in high and low density environments.  
      Bottom:  Axis ratio $b/a$ versus luminosity.  Symbols and 
      contours as in upper panels.  Small stars show the distribution
      of the BCG sample used in Bernardi et al. (2007a). }
 \label{mrzse}\label{ba}
 \end{center}
\end{figure*}

All the single galaxies in our sample were classified as being 
`round' if the SDSS axis ratio parameter $b/a>0.7$, and as being 
`elongated' otherwise.  This classification is consistent with a 
visual inspection of the isophotal structure in the ACS $i$-band 
image.  The top panels of Figure~\ref{ba} show the distribution 
of luminosities as a function of redshift; circles and diamonds 
show `round and `elongated' galaxies.  Notice that there are no 
round objects at low redshift.  What causes this?  

Vincent \& Ryden (2005) have analyzed the shapes of $z\le 0.12$ 
ellipticals in SDSS DR3, and find that the most luminous galaxies 
are rounder.  This is consistent with work prior to the SDSS 
(e.g. Tremblay \& Merritt 1996).  
The top left panel in Figure~\ref{mrzse} shows that the distribution 
of luminosities in our sample increases strongly with redshift (a 
consequence of the SDSS magnitude limit), and that the more 
luminous objects are indeed rounder.  

\begin{figure*}
 \begin{center}
 \includegraphics[scale=0.8]{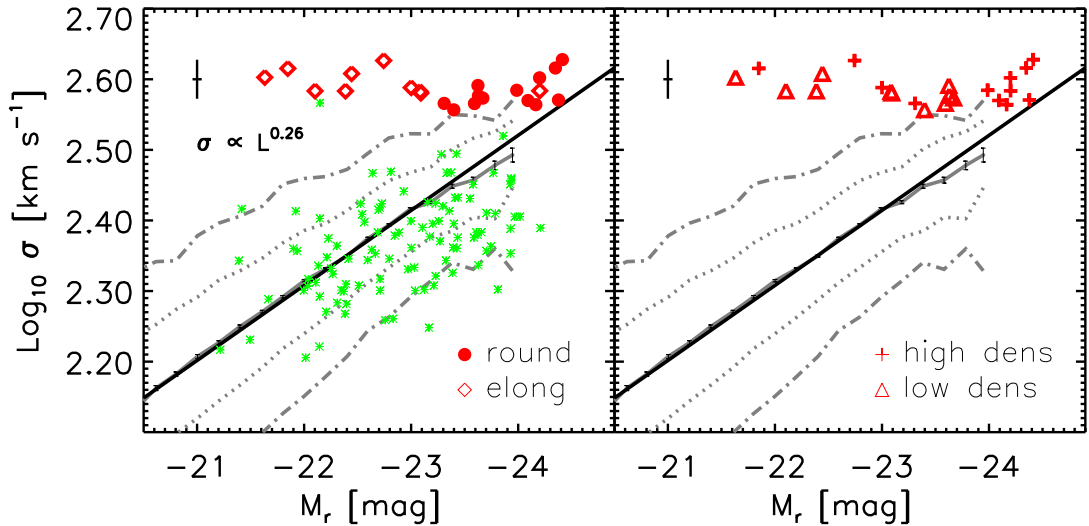}
 \caption{The velocity dispersion-luminosity relation.  
      Symbols with error bars show the median $\sigma$ in small 
      bins in $L$ for the bulk of the early-type population, 
      dotted and dot-dashed lines show the region populated by 68\% 
      and 95\% of the galaxies.  Thick solid line shows a linear 
      fit to this distribution to emphasize the flattening at 
      high-$\sigma$.  Small stars show the BCGs, and the other 
      symbols show our big-$\sigma$ sample.
      Filled circles and open diamonds in the panel on the left show 
      objects classified as `round' and `elongated';  
      triangles and crosses in the panel on the right indicate 
      low and high density environments. }
 \label{SL}
 \end{center}
\end{figure*}

To quantify if there is a real trend, we must compare the shapes in 
our sample with the expected shapes of galaxies of the same luminosity 
but less extreme velocity dispersions.  This is done in the bottom left 
panel of Figure~\ref{ba}.  Contours show the distribution of the bulk 
of the early-type population in the $b/a$-luminosity plane; the 
transition from solid to dotted lines delineates the region containing 
$68\%$ of the objects.  At luminosities fainter than $M_r=-22$ this 
distribution broadens, so the mean $b/a$ decreases -- this is 
responsible for most of the luminosity dependence we referred to 
previously.  Here, we are most interested in the brighter objects 
($M_r\la -22$) where, to lowest order, the distribution of $b/a$ 
appears to be almost independent of luminosity.  There is a hint 
of a decrease at the very bright end ($M_r\la -23.5$), a point to 
which we will return shortly.  

For comparison, stars show the shapes of the BCGs in the SDSS C4 
sample analyzed by Bernardi et al. (2007a); BCGs appear to have 
approximately the same distribution as the main galaxy sample -- 
but note that they too appear to have slightly smaller $b/a$ at 
$M_r<-23.5$.  
If we ignore this decrease, then the plot indicates that 
all but one of the big-$\sigma$ objects more luminous than 
$M_r=-23.1$ is `round', and all of the fainter objects are 
`elongated'.  In fact, the `round' objects in our sample tend 
to have values of $b/a$ which are about typical for their 
luminosity, whereas many of the others have lower than 
average $b/a$.  

The bottom panel on the right shows that, at fixed $L$, the big$-\sigma$ 
objects in high-density environments have larger $b/a$ 
(i.e., are rounder) than their counterparts in low-density 
environments, an interesting finding that we will not pursue 
further here.  

\begin{figure*}
 \begin{center}
 \includegraphics[scale=0.8]{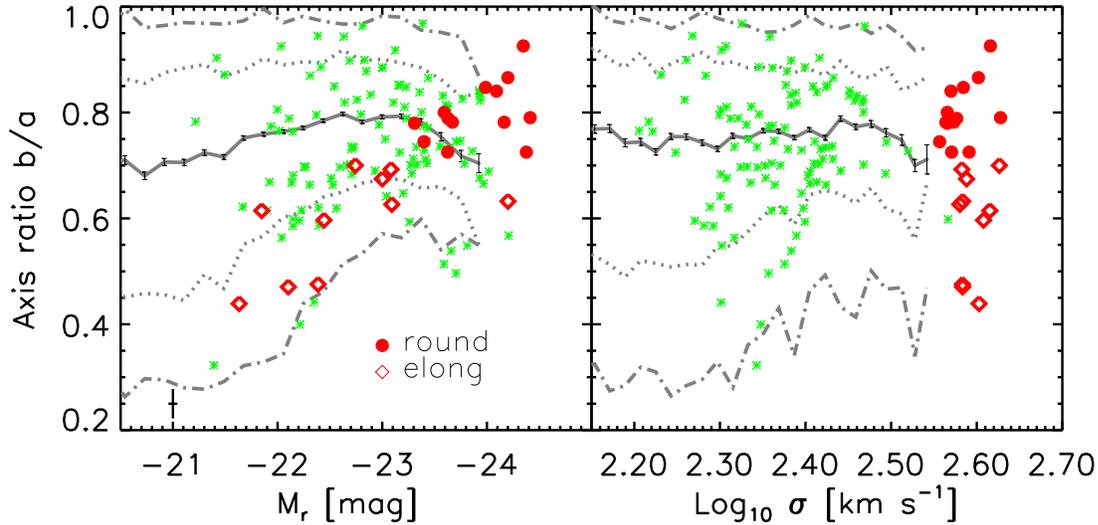}
 \caption{Shape parameter $b/a$ as a function of luminosity (left) 
      and velocity dispersion (right) for the bulk of the early-type 
      population; thick solid line shows the median, dotted lines 
      show the region populated by 68\% of the galaxies, and dot-dashed 
      lines show the same but for 95\%.  Stars show the distribution 
      of BCGs and circles and diamonds show our big-$\sigma$ sample. }
 \label{bazoom}
 \end{center}
\end{figure*}

\begin{figure*}
 \begin{center}
 \includegraphics[scale=0.8]{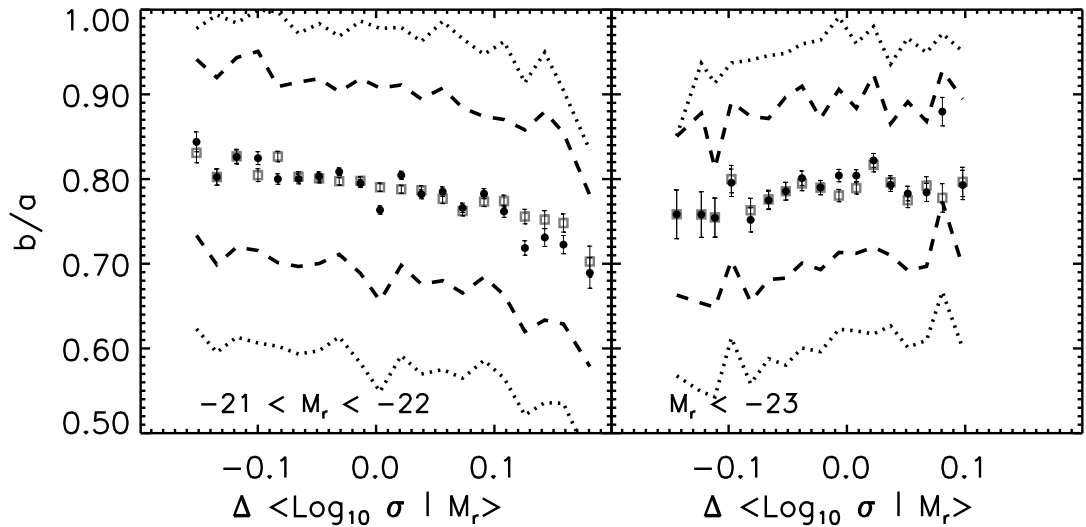}
 \caption{Correlation between axis ratio $b/a$ and residuals from 
          the mean $\sigma$ in two narrow bins in luminosity for the 
          bulk of the early-type galaxy population. Filled circles 
      show the median, dotted lines show the region populated by 68\% 
      of the galaxies, and dot-dashed lines show the same but for 95\%.
         At low luminosities objects which have small $\sigma$ for their 
          $L$ are slightly rounder; 
          at high luminosities, the objects which have large $\sigma$ 
          for their $L$ are rounder (they have large $b/a$).  Note 
          also that the scatter in $\sigma$ is smaller at large $L$. }
 \label{baLS}
 \end{center}
\end{figure*}

\subsection{Fast rotators at low luminosities?}\label{rotate?}
Figure~\ref{SL} shows the velocity dispersion-luminosity relation 
traced by these objects.  This illustrates that the high-luminosity 
objects outline the high-$\sigma$ boundary of the relation defined 
by the bulk of the population, whereas the low-luminosity objects 
are clearly extreme outliers.  These are primarily the `elongated' 
objects, for which projection effects cannot have caused the 
large velocity dispersions.  E.g., if the objects are prolate, 
then small $b/a$ means the long axis is perpendicular to the 
line of sight, so the velocity dispersion is not enhanced.  
If the objects are oblate, then small $b/a$ suggests they may be 
like thick disks viewed edge-on.  The SDSS velocity dispersion 
estimate comes from a single fiber, so it does not separate out 
the contribution to the observed velocity dispersion $\sigma$ 
which comes from ordered motions $v$.  
Hence, one might wonder if rotational motion $v$ has contributed to 
the velocity dispersion estimate of the low luminosity, big-$\sigma$ 
objects with small $b/a$.  

Although spatially resolved spectra would allow us to address 
this definitively, a more detailed study of the HST surface 
brightness profiles of these objects, in Hyde et al. (2008), 
is very suggestive.
Hyde et al. find that the most luminous objects tend to have 
shallower `core' inner profiles, whereas the less luminous objects 
tend to have `power-law' profiles with diskier isophotes.  
(We provide this core/power-law classification in 
Table~\ref{tab:classify}; in most cases, the difference between 
`core' and `power-law' is already apparent in the contour plots 
shown in Figures~\ref{A1}--\ref{A6}.)  
This dependence on luminosity is in good agreement with previous 
HST-based work on early-type galaxies (e.g. Faber et al. 1997; 
Laine et al. 2003; Ferrarese et al. 2006; Lauer et al. 2007).  
In the present context, it is significant that power-law objects 
tend to have higher rates of rotation (Faber et al. 1997).  

Moreover, there has been much recent interest in the distinction 
between slow rotators which tend to be luminous, and fast rotators 
which are not (e.g. Kormendy \& Bender 1996; Cappellari et al. 2007).  
The flatter galaxies in our sample ($b/a \le 0.6$) must have some 
degree of rotational support.  E.g., Binney (1978, 2005) suggests 
that $(v/\sigma)\approx (1-b/a)/(b/a)$.  Since the SDSS velocity 
dispersion estimate does not separate out the contribution from 
$v$ to the spectra, interpretation of the SDSS $\sigma$ for these 
objects is complicated.  We can get a rough idea of the size of 
the effect by noting that, for an isotropic oblate rotator with 
$b/a=0.6$, rotation adds 21\% to the total kinetic energy 
(Table~3 in Bender, Burstein \& Faber 1992); if $b/a=0.5$, this 
factor is 33\%.  If there is a disk, then this factor can be even 
larger.  


The SDSS spectra will be affected by this $v$ only if the light from 
the rotating component is a significant fraction of the total light 
in the fiber (also see Dressler \& Sandage 1983 for a discussion of 
the effect of rotation on the estimated $\sigma$).  
Hyde et al. (2008) also provide bulge-disk decompositions of these 
objects.  The elongated objects typically have bulge-to-total light 
ratios of order B/T$\approx 0.5$.  The half-light radii of the bulges 
are approximately equal to B/T times the values reported in Table~2 
(which are from single component fits).  If there is a disk, then 
it is likely that the bulge is also rotating, and this will contribute 
to the SDSS estimate of $\sigma$.  Moreover, for many of these objects, 
the angular half light radius of the bulge is slightly smaller than 
that of the SDSS fiber, so the dynamics of the disk may also have 
affected the spectrum.  Thus, it seems likely that some of the 
flattening of the less luminous objects in our sample is due to 
rotation, and that the velocity dispersion estimate has been enhanced 
because of this rotation.  


\begin{figure*}
 \begin{center}
 \includegraphics[scale=0.8]{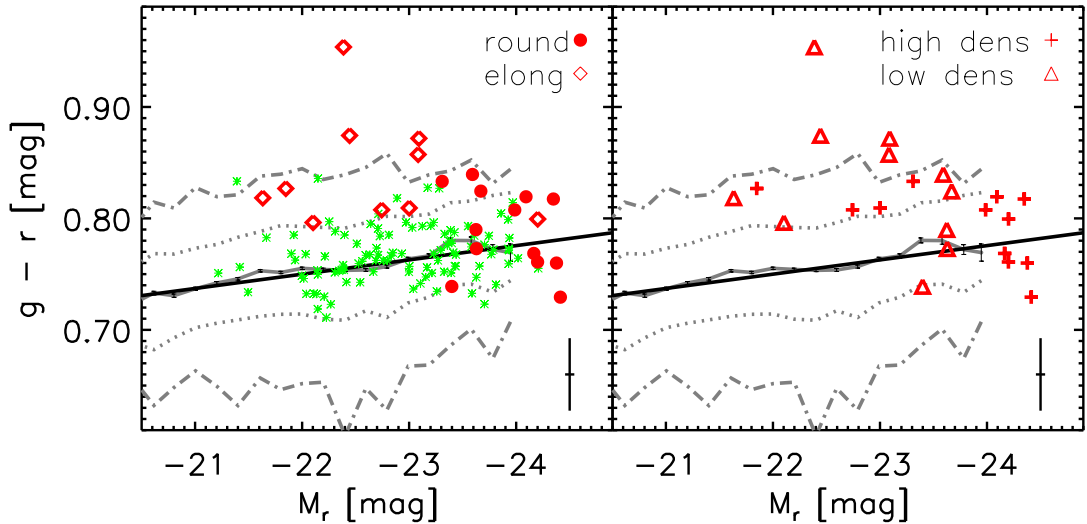}
 \caption{The color-magnitude relation of the bulk of the early-type 
      population (solid, dotted and dot-dashed lines), 
      the big-$\sigma$ sample (large symbols and dashed line) and BCGs (small stars).
      Line and symbol styles are as in Figure~\ref{SL}.
      The reddest object in the Figure has a dust-lane.}
 \label{CM}
 \end{center}
\end{figure*}

\subsection{Radial mergers at high luminosities?}\label{radial?}
With this possibility in mind, it is worth reconsidering the 
$b/a-$luminosity and $b/a-\sigma$ relations.  
The panel on the left of Figure~\ref{bazoom} shows the $b/a-$luminosity 
distribution, but now the solid curve shows the median $b/a$ for the 
bulk of the early-type population (defined for this figure as having 
$\sigma<350$~km~s$^{-1}$) in a number of narrow bins in luminosity.  
The dotted and dot-dashed curves indicate the regions containing 68\% 
and 95\% of the objects in each luminosity bin.  This shows that the 
gradual increase of $b/a$ with increasing $L$, that is consistent 
with previous work, reverses at $M_r\la -23.5$.  The BCGs appear to 
track this reasonably well (suggesting that it would be interesting 
to repeat the analysis of BCG shapes in Ryden et al. 1993, in which 
luminosity dependence was ignored).  If BCGs, or, more generally, 
the most luminous galaxies, formed from predominantly radial mergers, 
then one might expect them to be more prolate on average than the 
lower luminosity progenitors from which they formed.  Thus, the 
decrease in the mean $b/a$ at large luminosities may be indicating 
that the radial merger model is reasonable.  

The panel on the right shows a similar analysis of the $b/a-\sigma$ 
relation.  In this case, the bulk of the population (defined as 
having $\sigma<350$~km~s$^{-1}$) shows no trend, except for a 
slight decrease at $\log_{10}\sigma \ga 2.5$.  Although the four 
BCGs with the largest values of $\sigma$ lie below this relation, 
a larger sample is needed to conclude that this decrease in $b/a$ 
at large $\sigma$ is also seen in the BCG population.  

Notice that, in both panels, essentially all the `round' 
big-$\sigma$ objects lie above the mean relation for their 
luminosities or velocity dispersions, whereas all the `elongated' 
objects lie below it.  
If the velocity dispersions of the elongated objects have indeed 
been overestimated, then they should really be shifted to lower 
$\sigma$ (with no change in $L$).  And if $b/a$ is an indicator of 
$v/\sigma$, this shift is of order 0.08~dex (see Section~\ref{rotate?}).  
However, the  big-$\sigma$ objects which we classified as being 
`round' would not be shifted.  Because they are rounder than 
expected given the $b/a-L$ and $b/a -\sigma$ scalings for the bulk 
of the population, we cannot reject the hypothesis that these objects 
are prolate and viewed along the line of sight, and this has enhanced 
the observed velocity dispersions.  However, we argue below that 
this enhancement is unlikely to be more than a 10\% effect.  

If the velocity dispersions have been enhanced by projection, we 
might expect projection effects to contribute to the scatter in the 
$\sigma-L$ relation.  Figure~\ref{baLS} shows the correlation 
between axis ratio $b/a$ and residuals from the mean $\sigma$ in 
two narrow bins in luminosity for the bulk of the early-type galaxy 
population.  At low luminosities objects which have small $\sigma$ 
for their $L$ are slightly rounder (they have larger $b/a$).  This  
trend reverses at high luminosities, where the objects which have 
large $\sigma$ for their $L$ are rounder.  The sense of the scaling 
at large $L$ is expected in the projection model.  

\begin{figure*}
 \begin{center}
 \includegraphics[scale=0.7]{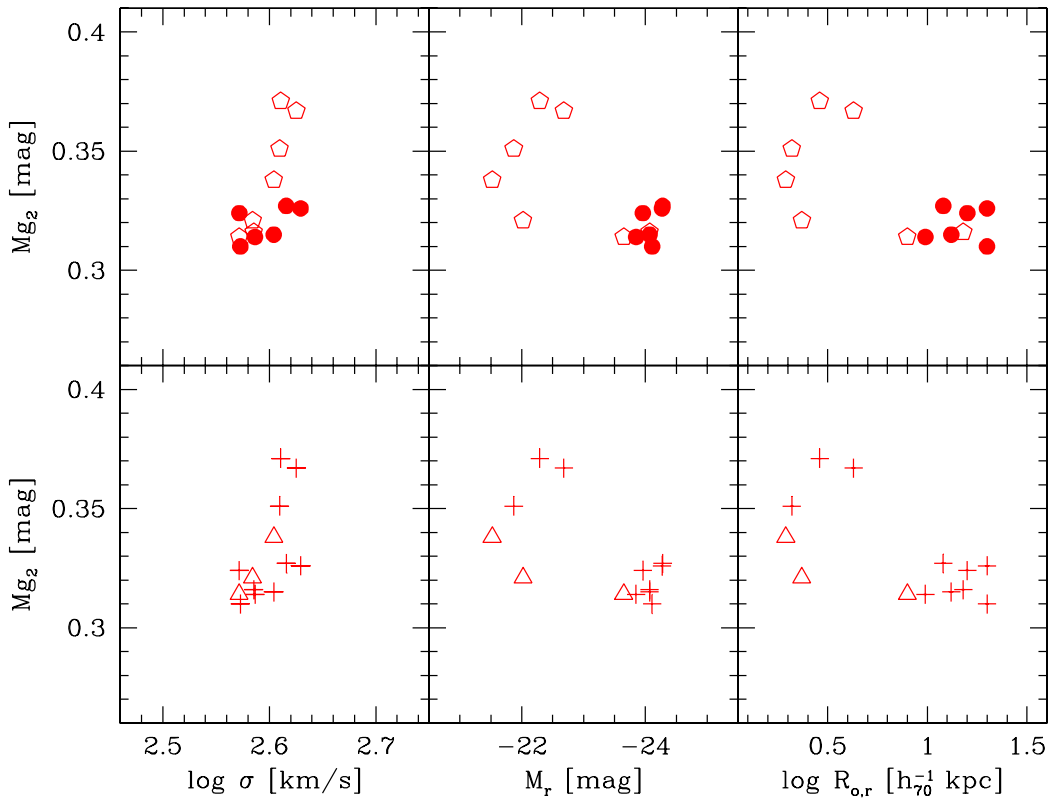}\vspace*{0.1cm}
 \caption{Mg$_2$ index as function of velocity dispersion (left), r-band
    luminosity (middle) and effective radius (right).  The flattened 
    low luminosity objects (open symbols in top panels) have 
    substantially larger Mg$_2$, so that Mg$_2$ is now anti-correlated 
    with luminosity and size.  
    Bottom panel shows that cluster galaxies tend to have larger Mg$_2$ 
    than more isolated galaxies of the same $L$. }
 \label{fig:mg2se}
 \end{center}
\end{figure*}

We show below that the mean $M_{\rm dyn}-L$ relation has small 
scatter, even at large $L$ (Figure~\ref{OvsL}).  If the long axis 
of a prolate object is perpendicular to the line of sight, then 
$R\propto\sqrt{ab}$, whereas $R\propto b$ if it lies along the line 
of sight.  If $\sigma$ were the same in both cases, then the two 
dynamical mass estimates would differ by a factor of $\sqrt{b/a} = 0.84$ 
($0.08$~dex) if $b/a=0.7$ (which Figure~\ref{bazoom} indicates is 
the mean value at $M_r<-23.5$).  This shows that projection effects 
could contribute substantially to the scatter in the 
$M_{\rm dyn}-L$ relation.  However, if the observed $\sigma$ is 
larger when viewed along the long axis (as one expects from the 
virial theorem), then the variation in the $M_{\rm dyn}$ estimate 
that is due to projection effects decreases 
(e.g. Gonz\'alez-Garc\'ia \& van Albada 2005).  
Since it is $\sigma^2$ which matters here, a $0.08$~dex 
variation in the size due to orientation effects could be cancelled 
by a $0.04$~dex variation in Log$_{10}$($\sigma$).  The scatter in 
$\sigma$ decreases at large $L$ (see Figure~\ref{SL}, or compare 
the range in the two panels of Figure~\ref{baLS}, or see Figure~3 
in Sheth et al. 2003); at $M_r<-23$ the observed rms scatter is 
0.05~dex, of which about 0.03~dex is not due to measurement errors.  
It is interesting that this is about the level expected if 
projection effects are beginning to dominate the scatter in the 
$\sigma-L$ relation.

\subsection{Colors, projection and dust} 
A possible problem with the projection model is that the most 
luminous objects in SAURON tend to be oblate, not prolate 
(Cappellari et al. 2007).  If our objects are oblate, then the 
round shapes we see ($b/a\ge 0.8$) suggest that we are seeing them 
face on, making the line of sight axis the shortest, rather than 
the longest one.  Therefore, as another test of the projection model 
for the `round' objects in our big-$\sigma$ sample, we have studied 
their colors.  

For the bulk of the early-type galaxy population, the color-magnitude 
relation is flat at fixed $\sigma$, with mean color increasing with 
$\sigma$ (Bernardi et al. 2005).  The filled circles in the left-hand 
panel of Figure~\ref{CM} show that, indeed, the round galaxies in 
our sample define a flat color-magnitude relation.  
If there were no dust, then one would not expect the colors to be 
affected by projection effects.
On the other hand, if these objects are dusty, prolate BCGs, viewed 
along the line of sight, one might expect them to appear redder 
than BCGs of the same $L$.  Figure~\ref{CM} suggests that they are 
not redder than BCGs, so if they are dusty, it will be difficult to 
accomodate the projection model.  One would then have to argue that 
they are intrinsically slightly bluer -- perhaps as a consequence 
of dry mergers of objects which were on similar positions in the 
color-magnitude relation.  The merger would increase the luminosity 
without changing the color, so moving them slightly brightward with 
respect to the color-magnitude relation.  At fixed luminosity, these 
objects would lie blueward of the mean relation if there were no 
dust; dust would bring them back onto the relation.  

In contrast to the `round' objects in our sample, about half of the 
`elongated' objects have extremely red colors.  However, the three 
reddest of these, SDSS J103344.2$+$043143.5, SDSS J112842.0+043221.7 
and SDSS J082646.7+495211.5, have obvious dust lanes.  
(One other object, SDSS J082216.5+481519.1, also has a dust lane; 
see Hyde et al. 2008 for further study of these objects.)  If 
these objects are fast rotators, then the presence of such dust lanes 
is not unexpected.  


\begin{figure*}
 \begin{center}
 \includegraphics[scale=0.8]{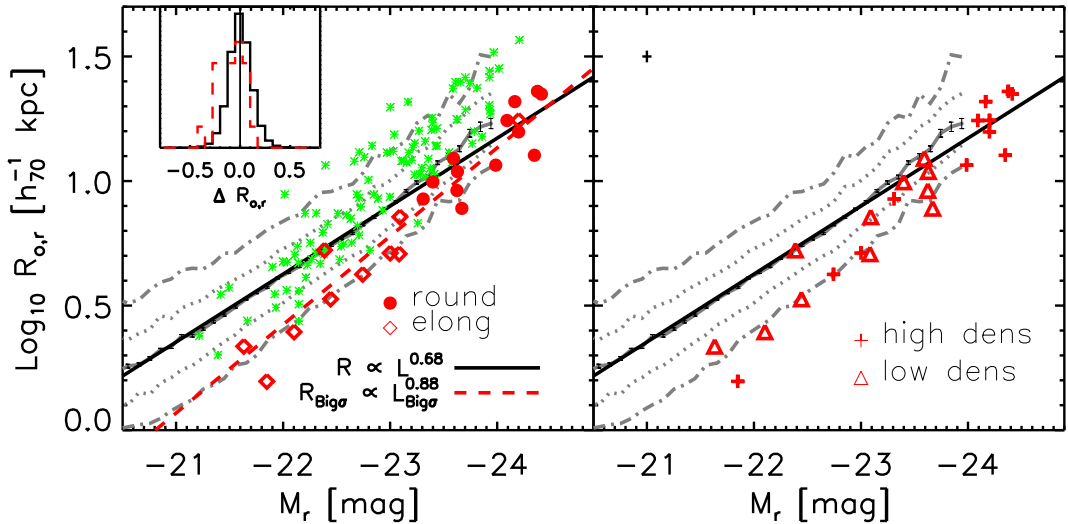}
 \caption{Correlation of size with luminosity for the bulk of the 
      early-type population (solid, dotted and dot-dashed lines) 
      and in our big-$\sigma$ sample 
      (symbols and dashed line).  
      Line and symbol styles are as in Figure~\ref{SL}.
      Stars show these correlations for BCGs.}
 \label{sizeL}
 \end{center}
\end{figure*}

Furthermore, the fast rotators seen by SAURON have anomolously large 
Mg$_2$ index strengths which is thought to be caused by stars which 
formed from metal-enriched gas.  Figure~\ref{fig:mg2se} shows that the 
flattened, low luminosity objects in our big$-\sigma$ sample (diamonds) 
also have abnormally large Mg$_2$ values; indeed Mg$_2$ is 
anti-correlated with luminosity and size.  
This suggests that, if rotation is important for these objects, 
then dissipation associated with the gastrophysics of star formation 
was also important.  Recall that these flattened objects 
tend to be the reddest objects for their luminosity - they are 
substantially redder than BCGs (Figure~\ref{CM}).  If Mg$_2$ is due 
to star formation, then it either happened long ago (else the colors 
would be bluer), or the redder colors are due to extreme metallicities 
or dust --  which we know is playing some role in at least half of 
these objects.  

\subsection{Brief summary}
To summarize this section, it appears that our sample of singles 
is made up of two populations.  The more luminous objects tend to 
be rounder and sit in crowded fields; it is possible that they 
are prolate objects, in most cases BCGs, viewed along the line of 
sight, in which case their velocity dispersions may be slightly 
enhanced compared to if they were oriented perpendicular to the 
line of sight.  The less luminous objects tend to be more flattened; 
if some of this flattening is due to rotation, then it may be that 
the velocity dispersion estimate has been enhanced because of this 
rotation.

\section{Scaling relations at large $\sigma$}\label{scaling}
We now study if the scaling relations of singles with large $\sigma$ 
are significantly different from those defined by the bulk of the 
early-type population.  

Figure~\ref{sizeL} shows that, at the high luminosity end, these 
objects (diamonds and circles) tend to be similar to the bulk of 
early-types (solid, dotted and dot-dashed lines), if one ignores 
the curvature in the size-luminosity relation (thick solid line).  
However, the solid line with error bars show that the $R-L$ relation 
curves upwards at high luminosities.  All our big-$\sigma$ objects 
lie below this curved relation.  In addition, they are systematically 
smaller than BCGs (stars) of the same luminosity, whatever the 
luminosity.  

\begin{figure*}
 \begin{center}
 \includegraphics[scale=0.8]{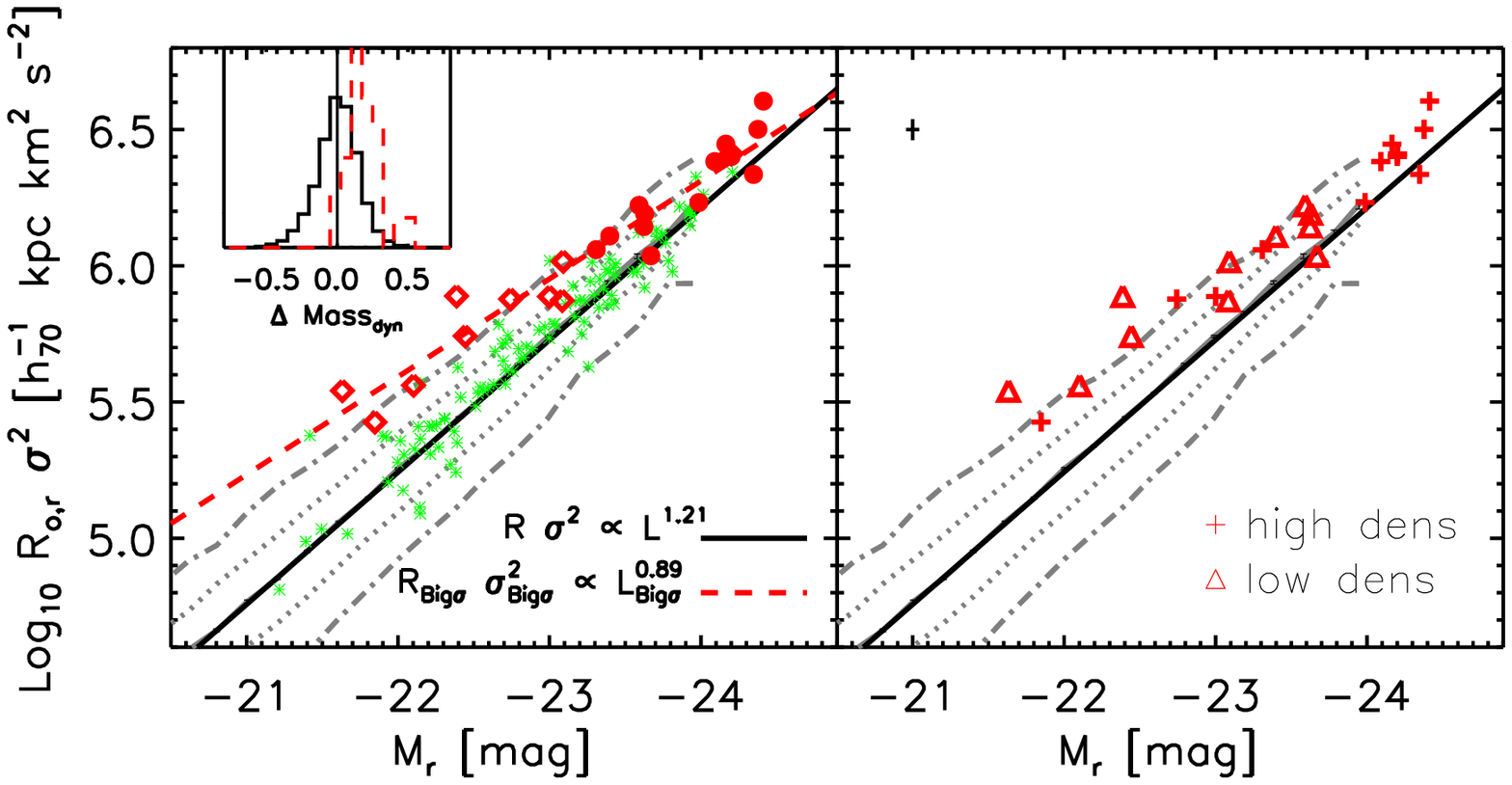}\vspace*{-0.5cm}
 \includegraphics[scale=0.8]{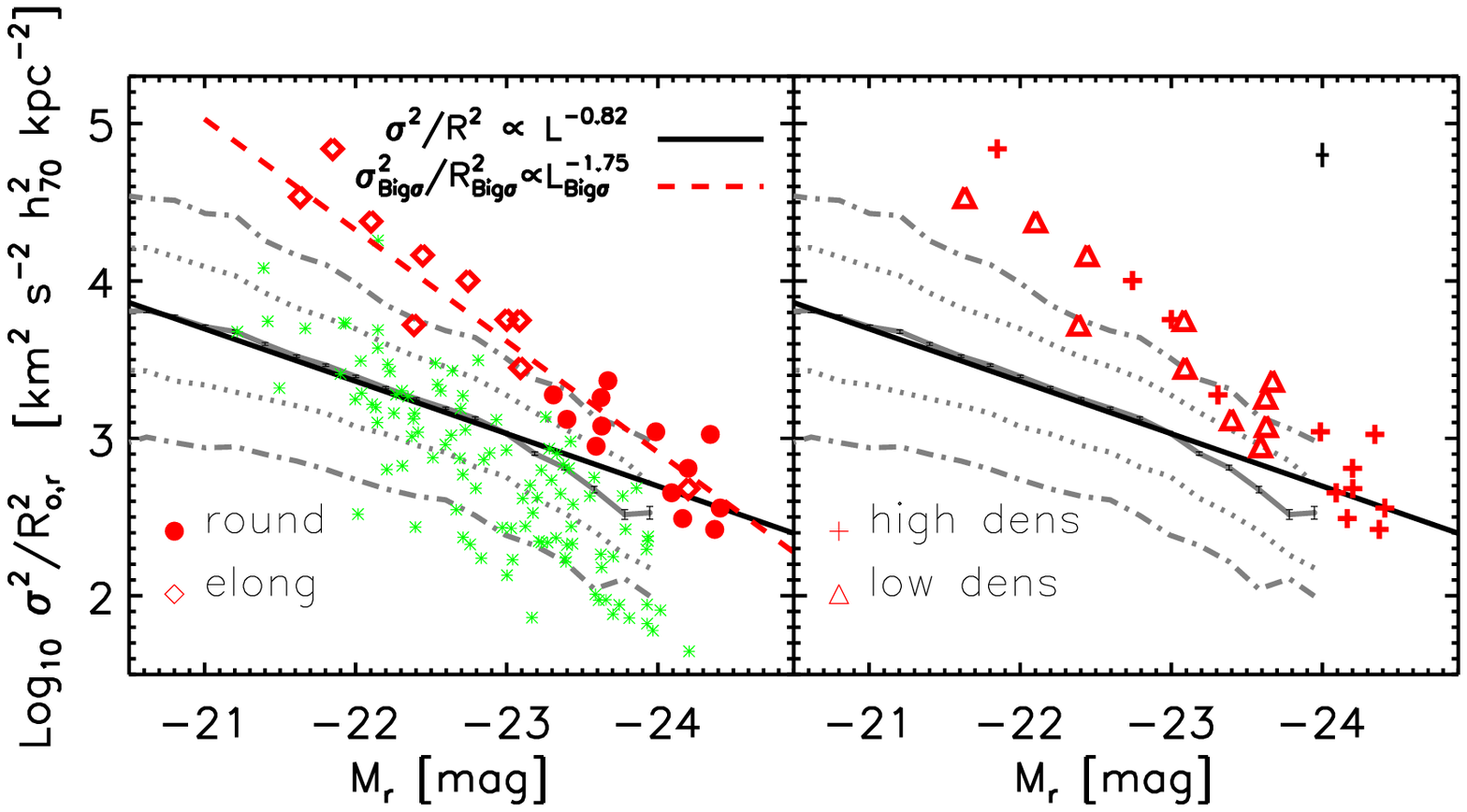}
 \caption{Correlation of dynamical mass (top) and density (bottom) 
      with luminosity for the bulk of the early-type population 
      (solid, dotted and dot-dashed lines) 
      and in the big-$\sigma$ sample (symbols and dashed line).  
      Line and symbol styles are as in Figure~\ref{SL}.  
      Stars show these correlations for BCGs. }
 \label{OvsL}
 \end{center}
\end{figure*}

The top panels of Figure~\ref{OvsL} show that these objects 
define a tight mass-luminosity relation (we set 
$M_{\rm dyn} \propto R\sigma^2$) 
that is offset from and has a different slope compared to that 
defined by the bulk of the population;  they are the most massive 
galaxies at any $L$, even compared to BCGs.  Only at the largest 
$L$ do BCGs have similar masses.  This shows again 
(c.f. Figure~\ref{bazoom}) that the most massive galaxies in the 
Universe have large $\sigma$ {\em and} $L$.  
We discuss the different slope and offset shortly.  

Since density $\propto (R\sigma^2/R^3)\propto (\sigma/R)^2$, the
large velocity dispersions and small sizes of these objects mean 
they are much denser than average galaxies of the same $L$.  
The bottom panels in Figure~\ref{OvsL} show that the density-$L$ 
relation they define is indeed very different from that of the bulk 
of the population.  Note in particular that they are typically about 
0.6~dex denser than BCGs of the same luminosity.  
At low luminosities, they appear to be more than ten times denser 
than average.  We will have more to say about this shortly.

The mass-luminosity and density-luminosity scalings are approximately 
consistent with the $R\propto L^{0.9}$ scaling, and the assumption 
that $\sigma$ is essentially constant, since this would predict 
 mass $\propto R\sigma^2 \propto L^{0.9}$, and 
 density $\propto (\sigma/R)^2 \propto L^{-1.8}$.  
Hence, to understand these other scalings, it is sufficient to 
understand why $R\propto L^{0.9}$, when the bulk of the population 
scales as $R\propto L^{0.68}$.  In this context, it is instructive 
to study the mean size at fixed $L$ {\em and} $\sigma$ in the bulk 
of the population.  

We have done this in two ways, one numerical, and the other analytic.
First, following arguments in Bernardi et al. (2005), 
\begin{equation}
 \frac{\langle R|M_r,V\rangle}{\sigma_R} =
  \frac{M_r}{\sigma_M} \frac{\rho_{RM} - \rho_{RV}\rho_{VM}}{1-\rho^2_{VM}} + 
  \frac{V}{\sigma_V} \frac{\rho_{RV} - \rho_{RM}\rho_{VM}}{1-\rho^2_{VM}},
\end{equation}
where $R$ and $V$ mean ${\rm log}(R/R_*)$ and 
${\rm log}(\sigma/\sigma_*)$, $M_r$ is absolute $r-$band 
magnitude minus $M_*$, $\sigma_x$ is the rms of the observable $x$, and 
$\rho_{xy}$ is the correlation coefficient of the observables $x$ and $y$.  
Hence, at fixed $\sigma$, the slope of the size-magnitude relation is 
\begin{equation}
 \frac{\delta \langle R|M_r,V\rangle}{\delta M_r}=\frac{\sigma_{R}}{\sigma_M}\,\frac{\rho_{RM} - \rho_{RV}\rho_{VM}}{1-\rho^2_{VM}}
 = \frac{-0.88}{-2.5},
\end{equation}
where we have inserted the values derived from the $R-L$, $\sigma-L$ 
and $R-\sigma$ relations defined by the early-type sample.  

\begin{figure*}
 \begin{center}
 \includegraphics[scale=0.8]{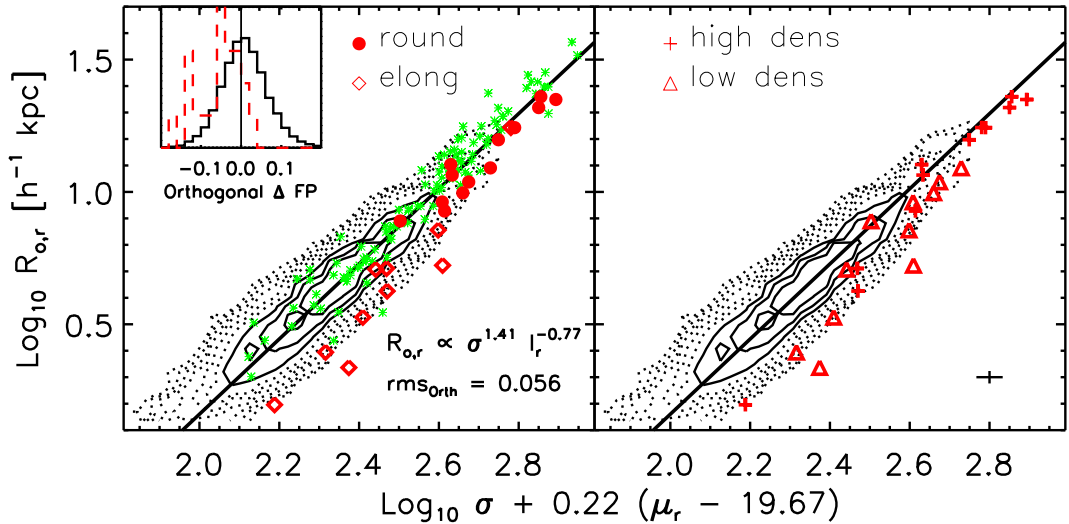}
 \caption{Location of the big$-\sigma$ sample (circles and diamonds) 
      with respect to the Fundamental Plane defined by the early-type 
      galaxy sample (contours); stars show the location of BCGs.
      Contours and symbols as in Figure~\ref{ba}.
 }
 \label{FP}
 \end{center}
\end{figure*}

Second, we restricted the full sample to narrow bins in $\sigma$ and 
measured the slope of the $R-L$ relation in the different subsamples.  
We found that the slope does not depend on the choice of bin.  
The result of shifting the zero-point to best-fit our big$-\sigma$ 
sample is shown as the dashed line in the panel on the left of 
Figure~\ref{sizeL}.  A similar analysis of the mass-$L$ and 
density-$L$ relations yields the dashed lines shown in 
Figures~\ref{OvsL}.  In all cases, the dashed lines provide a good 
description of our big-$\sigma$ sample.  

\begin{figure*}
 \begin{center}
 \includegraphics[scale=0.7]{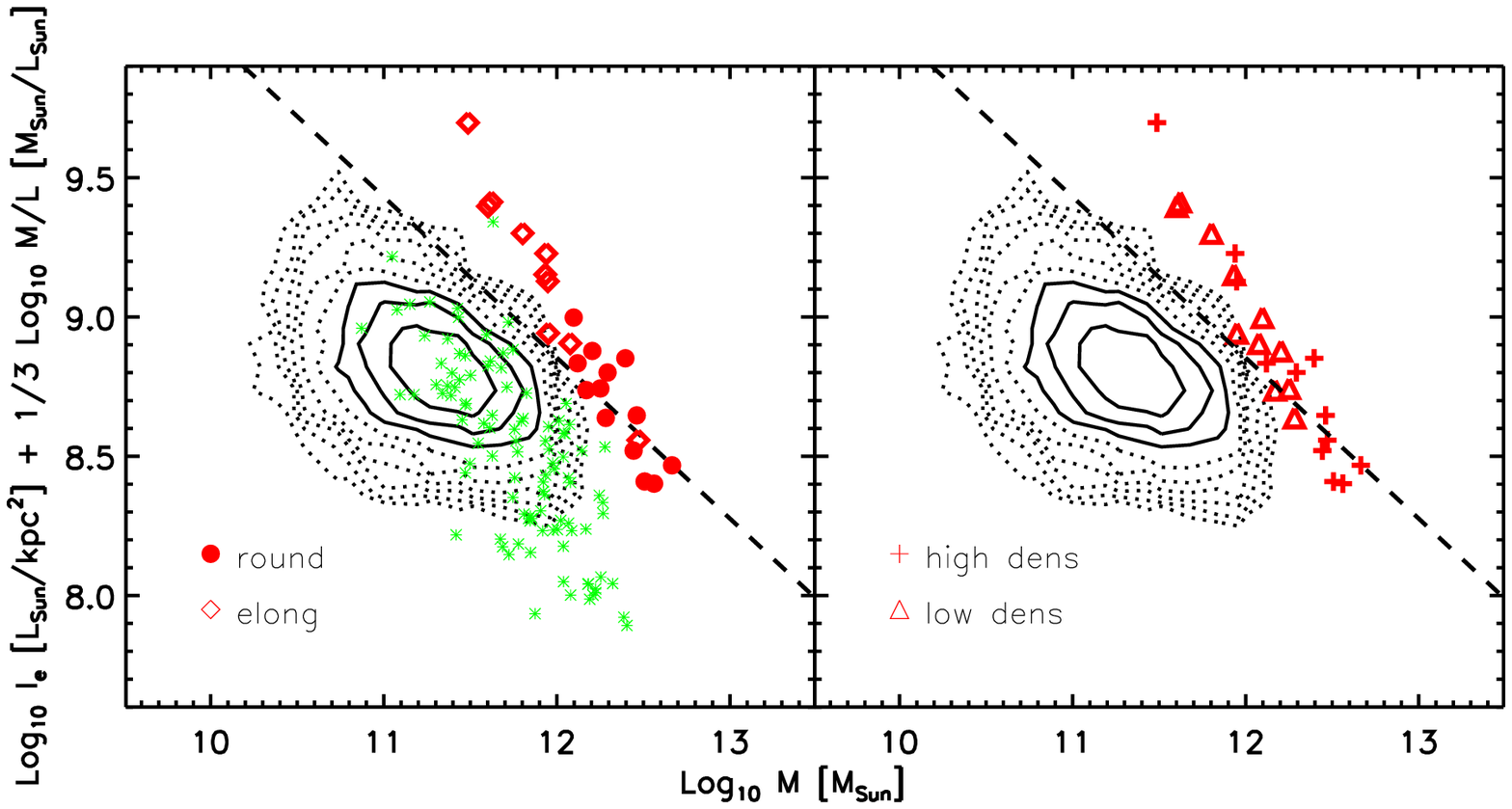}
 \includegraphics[scale=0.7]{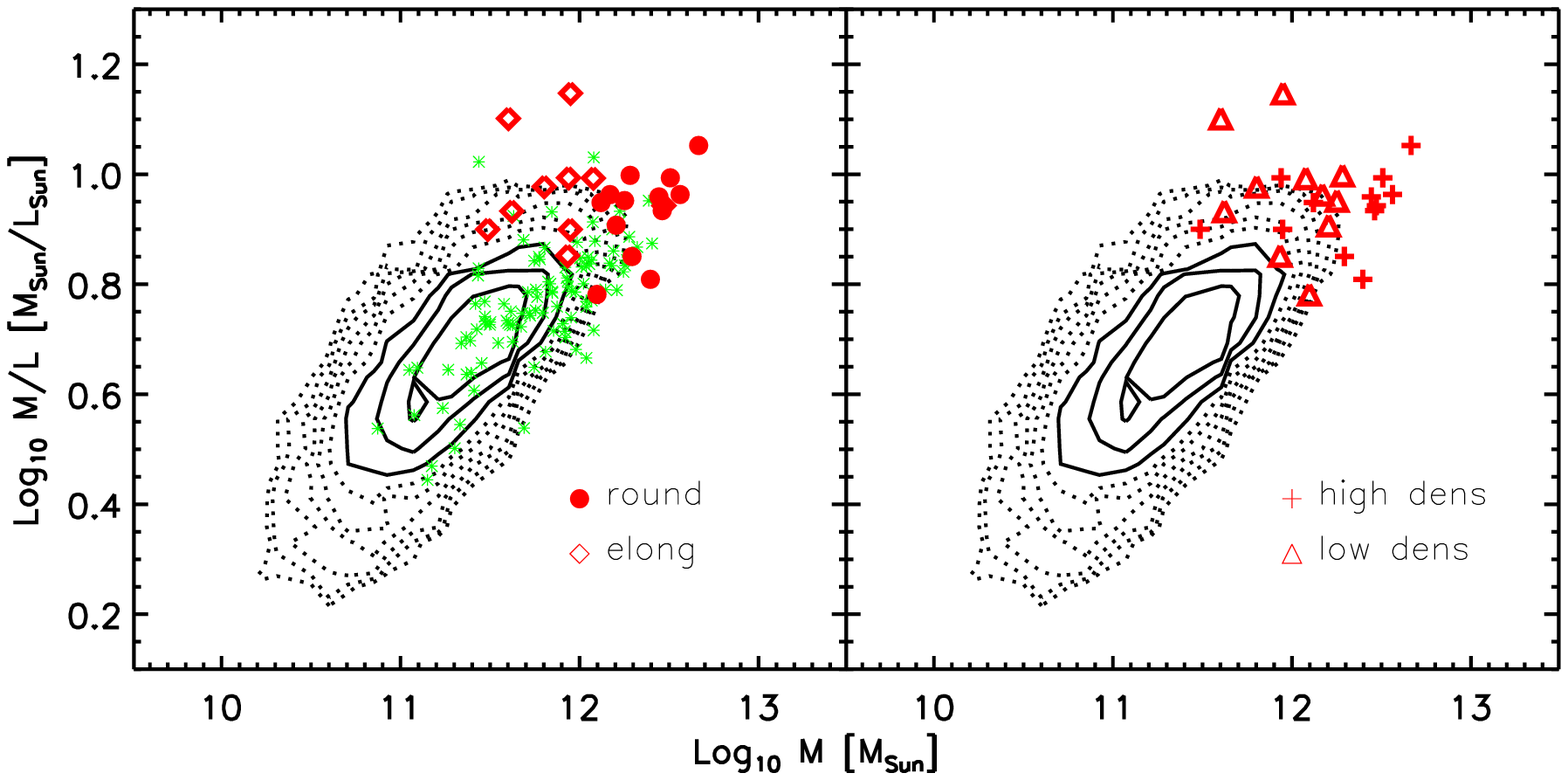}
 \caption{
      Distribution in $\kappa$-space (top panels: $\kappa_2$ versus 
      $\kappa_1$; bottom panels: $\kappa_3$ versus $\kappa_1$) of the 
      bulk of the early-types (contours), the big-$\sigma$ subsample, 
      and BCGs. Contours and symbols as in Figure~\ref{ba}.
      The figure shows $\kappa_1$, $\kappa_2$ and $\kappa_3$ transformed
      to solar units.
Dashed line sloping down and to right in top panels shows $\kappa_1+\kappa_2=8$
transformed to these variables.  
 }
 \label{kappa}
 \end{center}
\end{figure*}

Thus, both numerical and analytic arguments reproduce the steeper 
 $R\propto L^{0.9}$ scaling of our big-$\sigma$ sample.  
So it seems reasonable to conclude that the objects in our  
big$-\sigma$ sample are simply the large$-\sigma$ extremes of the 
early-type population -- they are not unusual in any other way 
(except, perhaps, their shapes).  
The underlying physical reason for this steeper relation can be 
understood as follows:  
If $R\sigma^2$ is linearly proportional to galaxy mass, then 
$(M/L)\, L = R \sigma^2$.  So, at fixed $\sigma$, we expect 
$R\propto (M/L)\, L$.  If the mass-to light-ratio were the same 
for all galaxies, then we would expect $R\propto L$, which is 
considerably steeper than the $R\propto L^{0.68}$ scaling which 
holds when one averages over all values of $\sigma$ in the 
population.  

It is interesting to contrast these scalings with the size-luminosity 
relation of BCGs.  The abnormally large sizes of BCGs suggest 
unusual formation histories - perhaps dominated by dry-mergers 
(e.g., Lauer et al. 2007; Bernardi et al. 2007a).  
So one might have wondered if the smaller sizes of the objects 
in our sample point to formation histories which are unusual in 
some other way.  The discussion above suggests that their 
distribution in the size-luminosity plane is no more unusual 
than one might have expected, given their unusually large 
velocity dispersions.  Since we already know they have large 
velocity dispersions -- that is how this sample was selected -- 
the size-luminosity relation contains no new information.  

Thinking of these objects as having fixed $\sigma$ also helps 
understand the location of these objects with respect to the 
Fundamental Plane defined by the bulk of the population 
(see Figure~\ref{FP}):  in essence, these objects trace out the 
size -- surface-brightness correlation, with zero-point set by 
the value of $\sigma$.  In this respect, the Fundamental Plane 
is not as useful a diagnostic of the properties of these galaxies as 
were the other scaling relations.  

Although the FP itself was not particularly informative, the 
$\kappa-$space projection (Bender, Burstein \& Faber 1992) is.  
The bottom panels in Figure~\ref{kappa} (the edge-on projection) 
show that the round objects in our sample have mass-to-light ratios 
which are similar to those of BCGs; they are slightly larger than 
those of the bulk of the population.  However, some of the elongated 
objects appear to have mass-to-light ratios that are even larger than 
those of BCGs.  
The face-on projection (top panels) shows that while the round 
objects in our sample lie close to the boundary of the 
zone-of-avoidance ($\kappa_1+\kappa_2\ge 8$) defined by the bulk 
of the population, the elongated objects lie well inside it.
Bender et al. associated this boundary with extreme dissipation, 
so Figure~\ref{kappa} suggests that the lowest mass, elongated 
objects in our big$-\sigma$ sample have undergone abnormally 
high amounts of dissipation.  

However, most of these abnormal objects have small $b/a$, so one 
might worry that this conclusion changes if they are indeed fast 
rotators.  In this case, the SDSS $\sigma$ is almost certainly an 
overestimate, making both the estimated mass and density larger 
than they really are.  Accounting for this will reduce the $M/L$ 
values and shift the position in the face-on view of $\kappa$-space, 
thus weakening the case for extreme dissipation.  To estimate the 
magnitude of this effect, we can use the bulge-disk decompositions 
in Hyde et al. (2008).  These suggest bulge sizes are typically 
about a factor of 2 smaller than reported here, whereas bulge 
luminosities are reduced by a smaller factor.  If we also correct 
the velocity dispersions downwards by about 20\% (see discussion 
in Section~\ref{rotate?}), then the net effect is to increase the 
density by 0.4~dex while making the magnitudes somewhat less than 
0.75~mags fainter.  Thus, the bulges will be closer to the 
$\kappa_1+\kappa_2=8$ boundary, but will remain the densest 
objects for their luminosities.  


\section{Discussion}\label{discuss}
We used HST imaging (Figures~\ref{A1}--\ref{A6}) to separate 
genuinely single objects from those which are superpositions in 
a sample drawn from the SDSS DR1 and chosen to have 
$\sigma\ge 350$~km~s$^{-1}$.   The abundance of these large 
$\sigma$ objects that are singles is consistent with that given 
by extrapolating fits to the SDSS velocity function to higher 
$\sigma$ -- there is no `toe' at $\sigma > 400$~km~s$^{-1}$ 
(Figure~\ref{phiv}).  

The scaling relations (size-$L$, mass-$L$ and density-$L$) defined 
by these objects are different from those defined by the bulk of 
the early-type galaxy population:  for a given luminosity, these 
objects are amongst the most massive and densest early-type galaxies 
(Figures~\ref{sizeL} and~\ref{OvsL}).  However, these differences can 
be understood by thinking of this sample as being the large-$\sigma$ 
tail of the early-type galaxy population, but not being otherwise 
unusual.  Table~2 lists the redshifts, luminosities, sizes, velocity 
dispersions, colors, Mg$_2$ abundances, shapes, and environments of 
these objects.  (The luminosities, sizes and shapes have been 
corrected for known problems with SDSS sky-subtraction; 
Section~\ref{photocorr}).  

These single galaxies with $\sigma\ge 350$~km~s$^{-1}$ appear to be 
of two types:  the more luminous objects ($M_r<-23.5$ or so) are 
round ($b/a> 0.7$), whereas the less luminous objects are flatter
(Figure~\ref{ba}).  In addition, Hyde et al. (2008) show that the 
HST-based surface brightness profiles of the low and high luminousity 
objects are `power-laws' and `cores' (in the language of 
Faber et al. 1997); the trend with luminosity is consistent with 
previous HST work on the centres of early-type galaxies, 
which shows that power-law galaxies may have significant rotation.  
The cores, in the galaxies which have them, are about a factor of 
ten smaller than the half-light radii.  However, the cores are not 
unusually small for the total $L$ or $\sigma$ (Hyde et al. 2008); 
this is in contrast to the half-light radii, which are (Figure~\ref{sizeL}).  
On the other hand, if the objects we classify as power-laws actually 
have cores that are below our resolution limit, then the upper 
limits we can set to the core-size are already at the low-end of 
the expected sizes for their $\sigma$, and perhaps also for their $L$ 
(Hyde et al. 2008).  

What do our observations imply for the formation histories of these 
two populations?  
At large $L$, these objects tend to be found in crowded fields so 
they may be BCGs.  If so, then it is likely that they formed from 
merging or accreting smaller galaxies, and this would explain why 
their inner profiles are shallow.  The puzzle is to explain why 
these objects are so much denser than the average galaxy or BCG 
of the same $L$ (bottom panels of Figures~\ref{OvsL} and~\ref{FP}), 
especially since the sizes of the inner core radii appear to be 
normal -- they do not appear to be small for their $L$ or $\sigma$ 
(Hyde et al. 2008).  
One possibility is that these objects formed from predominantly 
radial mergers with little angular momentum.  If the prolate objects 
which result are viewed along the long axis, this would produce 
slightly smaller half-light radii and slightly larger velocity 
dispersions.  

In this context, it is worth noting that the trend for the mean 
shape to become increasingly round as luminosity increases 
(e.g., Vincent \& Ryden 2005) appears to reverse at $M_r<-23.5$ 
(Figures~\ref{bazoom} and~\ref{baLS}).  This appears to be true 
for galaxies in the main sample, and for BCGs -- and is in 
qualitative agreement with models which postulate that radial 
mergers are common at large luminosities.  Compared to this 
decrease in $b/a$ at large $L$, the luminous objects in our 
big-$\sigma$ sample appear to be rounder than expected 
(Figure~\ref{bazoom}), consistent with the hypothesis that 
projection effects have resulted in smaller sizes and larger 
velocity dispersions.  
Nevertheless, even if one accounts for this effect, these objects are 
amongst the densest BCGs for their luminosities.  

Whereas projection effects may be important at large $L$, they almost 
certainly cannot account for the flattened shapes we see at low $L$.  
We can think of two plausible models for these flattened objects.  \\
i)  These are objects which contain a substantial component that 
    is rotationally supported.  \\
ii) These are objects in which gaseous dissipation has been most 
    efficient.\\
Option (i) must matter for the objects with $b/a\le 0.6$, since 
anisotropic dispersions cannot produce extremely flat shapes.  
Indeed, the fact that we see small $b/a$ only at small $L$ 
(Figure~\ref{ba}) suggests that these are examples of the low 
luminosity, fast-rotator population which has received considerable 
recent attention from e.g., the SAURON group (Cappellari et al. 2007).  
If so, then the SDSS velocity dispersions are artificially broadened 
by rotational motions, thus affecting the mass and density estimates.  
Reducing the mass estimates by the appropriate $b/a$-dependent factor 
would bring the flattened objects closer to the relation defined by 
the bulk of the population; it would also bring the objects closer to 
the $\kappa_1+\kappa_2=8$ boundary in $\kappa-$space (Figure~\ref{FP}).  

Further indirect evidence for rotation comes from Hyde et al. (2008) 
who show that these objects have `power-law' inner profiles, and 
significant amounts of dust.  Previous work has shown that such 
objects often have a significant rotational component 
(e.g. Laine et al. 2003).  
Determining if this is the case for our sample requires spatially 
resolved kinematics.  Nevertheless, we argue that even if one 
accounts for rotational motions, these objects are likely to remain 
the densest for their luminosities.  Thus, it appears that both (i) 
and (ii) are true for these objects (e.g., Figure~\ref{fig:mg2se} 
and related discussion).  

If the lower luminosity objects are indeed contaminated by rotation 
whereas the higher luminosity objects are not, then one might ask 
if the fact that both low and high luminosity objects define the 
same power-law scaling relations (Figures~\ref{sizeL}--\ref{kappa}) 
is simply fortuitous.  However, both rotational and random motions 
contribute to the kinetic energy in the virial theorem.  E.g., 
Appendix~B of Bender, Burstein \& Faber (1992) suggests that using 
only the true $\sigma$ of an isotropic oblate rotator underestimates 
the true mass by 35\% if $b/a=0.6$.  In addition, recent work on the 
velocity dispersion estimates of more distant objects suggests that 
including the effect of rotation on mass estimates may indeed be 
important (e.g. van der Vel \& van der Marel 2008).  So it may be 
that the contribution of ordered motions to SDSS velocity dispersion 
estimates helps to keep the scaling relations power-laws.  
In this regard, when comparing our sample of high-velocity dispersion 
galaxies with higher redshift z$\sim$1.5 samples of passive galaxies
 (Figure~19 in Cimatti et al. 2008),
the lower-luminosity galaxies in our sample populate a similar locus in
the size, mass, surface density plane as the superdense z$\sim$1.5
passive galaxies. It is possible that our low-redshift high-density galaxies
are the rare examples of the high-redshift superdense galaxies which
have not undergone any dry merging. This scenario is supported by the
fact that the low luminosity galaxies in our sample are in low-density
environments and have intact power-law centers. So it would be
interesting to check if the superdense z$\sim$1.5 galaxies are
``fast-rotators''.

It is common to predict black hole abundances by transforming an 
observed luminosity or velocity dispersion function using an assumed 
scaling relation between black hole mass $M_\bullet$ and 
galaxy $L$ or $\sigma$ (e.g. Lauer et al. 2007; Tundo et al. 2007; 
Shankar et al. 2008).  If one ignores the fact that there is 
scatter around the mean $M_\bullet-L$ or $M_\bullet-\sigma$ 
relations, then one might conclude that the big-$\sigma$ sample 
studied here would predict higher black hole masses from their 
$\sigma$ than from their $L$s.  However, the scatter is significant:  
the analysis in Bernardi et al. (2007b) shows how to include the 
possibility that the scatter in the $\sigma-L$ relation is correlated 
with scatter in the $M_\bullet-L$ and $M_\bullet-\sigma$ relations.  
See their Section 2.3 for a discussion of the effect of selecting 
objects with large $\sigma$ for their $L$.  

Finally, we note that although we have focussed on the single objects 
in this sample, the superpositions are interesting in their own right.  
Because they are close superpositions in both angle and redshift, in 
which the spectra show little or no sign of recent star formation, 
and because they provide information about smaller scales than is 
possible with ground based data, they can be combined with other 
HST-based samples of early-type galaxies (e.g. Laine et al. 2003; 
Lauer et al. 2007) to constrain dry-merger rates more precisely 
than previously possible (e.g. Bell et al. 2006; Masjedi et al. 2007; 
Wake et al. 2008).  
Such combined samples can also be used to build more realistic models 
of the expected configurations of multiple-lens systems.  
These studies are in progress.

\small
\begin{table*}
 \caption[]{Properties of the 23 objects identified as singles.  
            Superscript $c$ means the inner profile is shallow core 
            (from Hyde et al. 2008).  Magnitudes, sizes, color and $b/a$
            were computed by Hyde et al. (2008) on SDSS r-band images, 
            while velocity dispersions and Mg$_2$ index-strenghts are from 
            Bernardi et al. (2006).  }
\begin{tabular}{ccccccccccccccc}
 \hline &&&\\
  ID$_{\rm S}$ & $z$ & $M_r$ & $e_M$ & 
  $g-r$ & $e_{g-r}$ & log$_{10} R$ & $e_R$ &
  $\sigma$ & $e_\sigma$ & Mg$_2$ & $e_{{\rm Mg}_2}$ & $b/a$ & $e_{b/a}$  & Env \\
  & & [mag] & [mag] & [mag] & [mag] & [kpc] & [kpc] & kms$^{-1}$ & kms$^{-1}$ & [mag] & [mag] & & & \\
\hline &&&\\
 1 & 0.23949 & -23.40 & 0.06 & 0.74 & 0.04 &  1.00 &  0.03 & 360 &  37 &         -- &       -- &  0.74 &  0.04 & 0\\
 $2^c$ & 0.19827 & -23.31 & 0.04 & 0.84 & 0.03 &  0.93 &  0.02 & 367 &  28 &         -- &       -- &  0.78 &  0.03 & 1\\
 $3^c$ & 0.16773 & -23.59 & 0.03 & 0.84 & 0.02 &  1.09 &  0.02 & 367 &  29 &      0.307 &    0.010 &  0.80 &  0.02 & 0\\
 $4^c$ & 0.32787 & -24.16 & 0.06 & 0.78 & 0.05 &  1.32 &  0.03 & 366 &  52 &         -- &       -- &  0.78 &  0.04 & 1\\
 $5^c$ & 0.27775 & -24.09 & 0.05 & 0.83 & 0.04 &  1.24 &  0.03 & 371 &  30 &      0.324 &    0.009 &  0.84 &  0.04 & 1\\
 $6^c$ & 0.21356 & -23.67 & 0.04 & 0.83 & 0.03 &  0.89 &  0.02 & 374 &  22 &      0.314 &    0.008 &  0.78 &  0.02 & 0\\
 $7^c$ & 0.26271 & -24.38 & 0.04 & 0.77 & 0.04 &  1.36 &  0.02 & 372 &  29 &      0.310 &    0.009 &  0.73 &  0.03 & 1\\
 $8^c$ & 0.24639 & -23.63 & 0.05 & 0.78 & 0.04 &  1.04 &  0.03 & 377 &  33 &         -- &       -- &  0.79 &  0.03 & 0\\
 9 & 0.20533 & -23.09 & 0.06 & 0.88 & 0.04 &  0.86 &  0.03 & 380 &  34 &         -- &       -- &  0.63 &  0.03 & 0\\
10 & 0.22792 & -23.08 & 0.05 & 0.86 & 0.04 &  0.71 &  0.03 & 382 &  27 &         -- &       -- &  0.69 &  0.04 & 0\\
11 & 0.15939 & -22.39 & 0.04 & 0.96 & 0.03 &  0.72 &  0.02 & 383 &  41 &         -- &       -- &  0.48 &  0.02 & 0\\
12 & 0.15335 & -22.10 & 0.05 & 0.80 & 0.03 &  0.39 &  0.03 & 383 &  28 &      0.321 &    0.008 &  0.47 &  0.03 & 0\\
$13^c$ & 0.26275 & -24.20 & 0.04 & 0.81 & 0.03 &  1.24 &  0.02 & 383 &  34 &      0.316 &    0.009 &  0.63 &  0.02 & 1\\
$14^c$ & 0.23073 & -23.99 & 0.03 & 0.81 & 0.03 &  1.06 &  0.02 & 384 &  32 &      0.314 &    0.009 &  0.85 &  0.03 & 1\\
15 & 0.21930 & -23.00 & 0.06 & 0.81 & 0.04 &  0.71 &  0.03 & 387 &  41 &         -- &       -- &  0.67 &  0.04 & 1\\
16 & 0.28489 & -23.62 & 0.07 & 0.80 & 0.05 &  0.96 &  0.04 & 390 &  44 &         -- &       -- &  0.73 &  0.04 & 0\\
17 & 0.12705 & -21.63 & 0.05 & 0.82 & 0.03 &  0.34 &  0.02 & 400 &  28 &      0.338 &     0.010 &  0.44 &  0.02 & 0\\
$18^c$ & 0.26965 & -24.20 & 0.05 & 0.77 & 0.03 &  1.20 &  0.03 & 399 &  35 &      0.315 &    0.009 &  0.87 &  0.03 & 1\\
19 & 0.11610 & -21.85 & 0.03 & 0.83 & 0.02 &  0.20 &  0.02 & 412 &  27 &      0.351 &    0.009 &  0.61 &  0.02 & 1\\
20 & 0.16037 & -22.44 & 0.04 & 0.88 & 0.03 &  0.53 &  0.02 & 405 &  26 &      0.371 &    0.009 &  0.60 &  0.03 & 0\\
$21^c$ & 0.29718 & -24.35 & 0.04 & 0.82 & 0.04 &  1.10 &  0.02 & 412 &  27 &      0.327 &    0.007 &  0.93 &  0.03 & 1\\
$22^c$ & 0.13343 & -22.74 & 0.02 & 0.81 & 0.01 &  0.63 &  0.01 & 423 &  31 &      0.367 &     0.010 &  0.70 &  0.02 & 1\\
$23^c$ & 0.25026 & -24.41 & 0.04 & 0.74 & 0.03 &  1.35 &  0.02 & 424 &  30 &      0.326 &    0.006 &  0.79 &  0.02 & 1\\
\hline &&&\\
\label{tab:singles} 
\end{tabular}
\end{table*}
\normalsize





\section*{Acknowledgments}
We would like to thank the referee for suggestions which improved the paper.
We thank the HST support staff during Cycles 13 and 14 during which 
the SNAP-10199 and SNAP-10488 programs were carried out.
Support for programs SNAP-10199 and SNAP-10488 was provided through 
a grant from the Space Telescope Science Institute, which is operated 
by the Association of Universities for Research in Astronomy, Inc., 
under NASA contract NAS5-26555. 
J. H., A.F. and M.B. are grateful for additional support provided by 
NASA grant LTSA-NNG06GC19G. 

Funding for the Sloan Digital Sky Survey (SDSS) and SDSS-II Archive has been
provided by the Alfred P. Sloan Foundation, the Participating Institutions, the
National Science Foundation, the U.S. Department of Energy, the National
Aeronautics and Space Administration, the Japanese Monbukagakusho, and the Max
Planck Society, and the Higher Education Funding Council for England. The
SDSS Web site is http://www.sdss.org/.

The SDSS is managed by the Astrophysical Research Consortium (ARC) for the
Participating Institutions. The Participating Institutions are the American
Museum of Natural History, Astrophysical Institute Potsdam, University of Basel,
University of Cambridge, Case Western Reserve University, The University of
Chicago, Drexel University, Fermilab, the Institute for Advanced Study, the
Japan Participation Group, The Johns Hopkins University, the Joint Institute
for Nuclear Astrophysics, the Kavli Institute for Particle Astrophysics and
Cosmology, the Korean Scientist Group, the Chinese Academy of Sciences (LAMOST),
Los Alamos National Laboratory, the Max-Planck-Institute for Astronomy (MPIA),
the Max-Planck-Institute for Astrophysics (MPA), New Mexico State University,
Ohio State University, University of Pittsburgh, University of Portsmouth,
Princeton University, the United States Naval Observatory, and the University
of Washington.



{}


\label{lastpage}


\begin{thebibliography}{}

\bibitem[]{} Abazajian, K., et al. 2003, AJ, 126, 2081 

\bibitem[]{} Abbas, U. \& Sheth, R. K. 2007, MNRAS, 378, 641

\bibitem[]{} Almeida, C., Baugh, C. M. \& Lacey, C. G., 2007, MNRAS, 376, 1711

\bibitem[]{} Bell E., et al., 2006, ApJ, 652, 270

\bibitem[]{} Bender, R., Burstein, D. \& Faber, S. M. 1992, ApJ, 399, 462 

\bibitem[]{} Bernardi, M., Sheth, R. K., Nichol, R. C. et al. 2005, AJ, 129, 61

\bibitem[Bernardi et al.(2006)]{} 
Bernardi, M., Sheth, R.~K., Nichol, R.~C., Miller, C.~J.,
Schlegel, D., Frieman, J., Schneider, D.~P., Subbarao, M.,
York, D.~G., Brinkmann, J. 2006, AJ, 131, 2018

\bibitem[Bernardi et al.(2007)]{}
Bernardi, M., Hyde J.~B., Sheth, R.~K., Miller, C.~J.,
Nichol, R.~C. 2007a, AJ, 133, 1741

\bibitem[Bernardi et al.(2007b)]{}
Bernardi, M., Sheth, R.~K., Tundo, E, \& Hyde, J.~B. 2007b, ApJ, 660, 267

\bibitem[]{} Bernardi, M. 2007, AJ, 133, 1954

\bibitem[]{} Binney, J. 1978, MNRAS, 183, 501

\bibitem[]{} Binney, J. 2005, MNRAS, 363, 937

\bibitem[]{} Boylan-Kolchin M., Ma C.-P., Quataert E., 2006, MNRAS, 369, 1081

\bibitem[]{} Cappellari, M., Emsellem, E., Bacon, R. et al. 2007, MNRAS, 379, 418

\bibitem[]{} Cimatti, A., et al.\ 2008, A\&A, 482, 21

\bibitem[]{} Crawford C. S., Allen S. W., Ebeling H., Edge A. C., 
             Fabian A. C., 1999, MNRAS, 306, 857

\bibitem[]{} De Lucia, G., Springel, V., White, S. D. M., Croton, D. \& Kauffmann, G. 2006, MNRAS, 366, 499

\bibitem[]{} Dressler, A. \& Sandage, A. 1983, 265, 664

\bibitem[]{} Faber, S.~M., et al. 1997, aJ, 114, 1771 

\bibitem[]{} Ferrarese L. \& Merritt D., 2000, ApJ, 539, L9

\bibitem[]{} Ferrarese, L., Cote, P., Blakeslee, J.~P., Mei, S., Merritt, D., 
\& West, M.~J.\ 2006, submitted, arXiv:astro-ph/0612139 

\bibitem[Fruchter \& Hook (2002)]{FH02}
Fruchter, A.~S., \& Hook, R.~N. 2002, PASP, 114, 144

\bibitem[]{} Gebhardt K. et al. 2000, ApJL, 539, 16

\bibitem[]{} Gonz\'alez-Garc\'ia, A. C. \& van Albada, T. S. 2005, MNRAS, 361, 1043 

\bibitem[]{} Hoessel, J. G., Oegerle, W. R. \& Schneider, D. P. 
 1987, AJ, 94, 1111 

\bibitem[]{} Hyde, J., Bernardi, M., Fritz, A., Gebhardt, K., 
 Nichol, R. C. \& Sheth, R. K., 2008, MNRAS, submitted

\bibitem[]{} Hyde, J. \& Bernardi, M. 2008, MNRAS, submitted

\bibitem[]{} Kormendy, J. \& Bender, R. 1996, ApJ, 459, 57

\bibitem[]{} Laine, S., van der Marel, R.~P., Lauer, T.~R., Postman, M., O'Dea, C.~P., \& Owen, F.~N.\ 2003, AJ, 125, 478 

\bibitem[]{} Lauer, T.~R., et al. 2007, ApJ, 662, 808

\bibitem[]{} Malumuth E.~M. \& Kirshner, R.~P. 1981, ApJ, 251, 508

\bibitem[]{} ---. 1985, ApJ, 291, 8

\bibitem[]{} Masjedi M., Hogg D. W., Blanton M. R., 2007, ApJ, submitted, 
             (arXiv:0708.3240)

\bibitem[]{} Mo, H. J. \& White, S. D. M. 1996, MNRAS, 282, 347

\bibitem[]{} Oegerle W. R. \& Hoessel J. G. 1991, ApJ, 375, 15

\bibitem[]{} Postman M. \& Lauer T. R., 1995, ApJ, 440, 28
 
\bibitem[]{} Ryden, B. S., Lauer, T. R. \& Postman, M. 1993, ApJ, 410, 515 

\bibitem[]{as} Sandage A., 1976, ApJ, 205, 6 

\bibitem[]{} Scott E., 1957, AJ, 62, 248

\bibitem[]{} Schombert J.~M. 1987, ApJS, 64, 643

\bibitem[]{} ---. 1988, ApJS, 328, 475

\bibitem[]{} Schweizer F. 1982, ApJ, 252, 455

\bibitem[]{} Shankar, F., Weinberg, D. H., \& Miralda-Escud\'{e}, J. 2008, ApJ, submitted, arXiv/0710.4488 

\bibitem[]{} Sheth R. K., Tormen G. 2002, MNRAS, 329, 61

\bibitem[]{} Sheth, R. K., Bernardi, M., Schechter, P. L., et al. 
 2003, ApJ, 594, 225

\bibitem[]{} Thuan T.~X. \& Romanishin W. 1981, ApJ, 248, 439

\bibitem[Tremblay \& Merritt(1996)]{TM96}
 Tremblay, B., \& Merritt, D. 1996, AJ, 111, 2243

\bibitem[]{} Tundo E., Bernardi M., Hyde J. B., Sheth R. K., Pizzella A.
 2007, ApJ, 663, 53

\bibitem[]{} van der Vel, A. \& van der Marel, R. 2008, ApJ, in press, arXiv/0804.4228 

\bibitem[]{} Vincent R. A., Ryden B. S. 2005, ApJ, 623, 137

\bibitem[]{2slaq} Wake D. et al. 2008, MNRAS, 387, 1045

\end{thebibliography}
\end{document}